\documentclass[pdflatex,sn-mathphys-num]{sn-jnl}

\usepackage{amsmath,amssymb,amsthm,mathtools,bm}

\usepackage{hyperref}
\usepackage{cleveref}
\usepackage{algorithm}
\usepackage{algpseudocode}
\usepackage{booktabs}
\usepackage{graphicx}
\usepackage{xcolor}
\usepackage{enumitem}
\usepackage{thmtools}

\hypersetup{colorlinks=true,linkcolor=blue!50!black,citecolor=blue!50!black,urlcolor=blue!50!black}
\setlist{nosep}
\usepackage{amsmath,amssymb,amsthm}
\usepackage{mathtools}
\usepackage{geometry}
\usepackage{hyperref}
\usepackage{graphicx}
\usepackage{booktabs}
\usepackage{array}
\usepackage{mathtools}
\usepackage{algorithm, algpseudocode}
\usepackage{bm}
\usepackage{xcolor}
\usepackage{enumitem}
\usepackage{thmtools}
\usepackage{cleveref}

\newcommand{\R}{\mathbb{R}}
\newcommand{\X}{\mathcal{X}}
\newcommand{\U}{\mathcal{U}}
\newcommand{\W}{\mathcal{W}}
\newcommand{\Z}{\mathcal{Z}}

\newcommand{\pd}[2]{\dfrac{\partial #1}{\partial #2}}

\providecommand{\Zt}{\mathcal{Z}_t}
\newcommand{\Lfh}[1]{\mathcal{L}_{\hat f}^{#1}}
\newcommand{\Lgh}[1]{\mathcal{L}_{\hat g_{#1}}}
\newcommand{\V}{\mathcal{V}}
\newcommand{\E}{\mathcal{E}}
\newcommand{\D}{\mathcal{D}}

\newcommand{\umax}{u_{\max}}

\DeclareMathOperator{\diag}{diag}

\theoremstyle{plain}
\newtheorem{theorem}{Theorem}
\newtheorem{lemma}{Lemma}
\newtheorem{proposition}{Proposition}
\newtheorem{corollary}{Corollary}
\theoremstyle{definition}
\newtheorem{definition}{Definition}
\newtheorem{assumption}{A}

\theoremstyle{remark}
\newtheorem{remark}{Remark}
\crefname{assumption}{Assumption}{Assumptions}
\crefname{theorem}{Theorem}{Theorems}
\crefname{lemma}{Lemma}{Lemmas}
\crefname{corollary}{Corollary}{Corollaries}
\crefname{proposition}{Proposition}{Propositions}
\crefname{remark}{Remark}{Remarks}
\crefname{algorithm}{Algorithm}{Algorithms}

\begin{document}
	
	\title[Data-Driven Sparse Identification and Robust Tube-Based Predictive Control for Multivariable Chaotic Systems]{Data-Driven Sparse Identification and Robust Tube-Based Predictive Control for Multivariable Chaotic Systems}
	
	\author[1]{\fnm{Shahin} \sur{Razani}}\email{s.razani@email.kntu.ac.ir}
	
	\affil[1]{\orgdiv{Faculty of Electrical Engineering}, \orgname{K.N. Toosi University of Technology}, \orgaddress{\city{Tehran}, \country{Iran}}}

	\abstract{	This paper addresses the data-driven synchronization and robust control of multivariable chaotic systems subject to strict actuation constraints, modeling uncertainties and bounded input disturbances. Using empirical data collected from a four-dimensional chaotic R\"ossler response system, Sequentially Thresholded Least Squares (STLSQ) sparse regression is employed to recover a nominal control-affine model, achieving an identification accuracy of $R^2 = 1.0000$ across all channels with root-mean-square error within $[4.15\times 10^{-6}, 8.33\times 10^{-4}]$. To establish a mathematically rigorous bridge between data-driven identification and robust synthesis, all structural discrepancies in the drift vector field and input matrix are unified into a single lumped uncertainty field $\delta(x,u)$. A computable scalar bound $\bar{\delta}$ is established based on library truncation error, regression residuals, input bounds, and Lipschitz constants (Lemma~1). We subsequently prove that input--output feedback linearization maps $\delta(x,u)$ into a state- and control-dependent additive disturbance $d(z,\nu)$ in Brunovsky canonical coordinates, bounded by $\|d(z,\nu)\|_2 \le \alpha_1 \bar{\delta} + \alpha_2 \bar{\delta} \|\nu\|_2$ with closed-form expressions for $\alpha_1$ and $\alpha_2$ (Theorem~1). This certified bound, together with a bounded input disturbance and a zero-order-hold margin, yields a compact disturbance set that seeds a robust tube-based model predictive controller. We prove that the resulting tightened quadratic program is recursively feasible, that the physical actuator limit ($|u_i| \le 1100$) remains inactive under an explicit sufficient condition, and that the closed-loop tracking error is Input-to-State Stable (ISS) with an asymptotic gain that is linear in $\bar{\delta}$ plus a term due to the exogenous input disturbance (Theorems~2 and~3).}
	
	
	\maketitle
	\section{Introduction}

\subsection{Motivation}

Chaotic oscillators are attractive building blocks for secure communication, encryption and sensing, and every such application rests on one capability: making a response system follow a drive system whose trajectory is aperiodic and exponentially sensitive to initial conditions~\cite{xiao2025preset,babanli2024synchronization,saeed2023new}. In practice the problem is rarely posed in its textbook form. The actuator saturates, the input is corrupted by disturbances, the controller runs at a finite sampling rate, and the equations of the response system are known only through measurements. Each of these departures is manageable on its own. Together they make a controller that was designed on paper behave differently once it runs on hardware, and a guarantee derived for the nominal model says little about the system that was actually built.

We study this situation on a deliberately demanding testbed, a four-dimensional hyperchaotic R\"ossler drive--response pair, in which the response is driven through two input channels limited to $\pm1100$ and perturbed by a bounded disturbance. The question is not only whether synchronization can be achieved, but what can be certified about it when the model of the response has been learned from a single short record. Our answer combines three ingredients that are usually treated separately: sparse regression to recover the model, input--output feedback linearization to reduce the synchronization error to a linear system, and tube-based predictive control to enforce the constraints. The contribution of the paper is the bridge between them.

\subsection{Related work}

\paragraph{Synchronization of chaotic systems.}
The synchronization literature is large and mature. Classical constructions rely on Lyapunov functions or on conditional Lyapunov exponents. Yao et al.\ propose a different mechanism, in which two systems synchronize when their energy variation rates agree~\cite{yao2024energy}, and Butusov et al.\ show that a time-reversible integration scheme generalizes the Pecora--Carroll approach and synchronizes systems from sparse or noisy data of a single state variable without any controller, although the outcome depends strongly on the initial conditions~\cite{butusov2025fast}. Controller-based schemes cover a wide range of settings: generalized synchronization of integer-order, fractional-order and delayed systems with feedback and finite-time controllers~\cite{shukla2025generalized}, adaptive laws derived from Lyapunov's second method for a new three-dimensional chaotic system~\cite{saeed2023new}, adaptive sliding mode control with parameter identification for multi-wing attractors~\cite{wu2025adaptive}, and control and synchronization of a discrete fractional-order model~\cite{kahouli2025chaos}. Intelligent controllers have joined this toolbox, among them a preset-time fuzzy zeroing neural network validated on an FPGA~\cite{xiao2025preset} and a fuzzy Z-number controller for secure communication~\cite{babanli2024synchronization}. These works are mainly concerned with the convergence of the synchronization error under a known model structure, or one that is uncertain only through a few parameters. Hard actuation limits and a quantitative description of the disturbance are not the central object of their analysis, and that is precisely where the present paper starts.

\paragraph{Identification of chaotic dynamics from data.}
When the model is not available, identification enters the picture. Peng and Li identify the parameters of the Lorenz system with observer/Kalman-filter and bilinear-transform schemes and use the estimate to initialize an adaptive synchronization law; they also observe that extremely small parameter deviations can produce dramatic mismatches in the predicted output of a chaotic model~\cite{peng2022parameters}. This observation is the core difficulty of any data-driven design. For a chaotic system, trajectory-level accuracy is a poor proxy for the quality of a model, and what a controller needs is a bound on the error of the vector field. Learning-based methods have widened the class of systems that can be modelled. Wu et al.\ estimate the parameters of discrete fractional chaotic systems with neural networks and report robustness to noise~\cite{wu2024data}. Reduced-order models built from autoencoders and neural ordinary differential equations capture the spatiotemporal chaos of the Kuramoto--Sivashinsky equation and have been used to train a reinforcement-learning controller~\cite{linot2022data,zeng2022data}. Spectral-submanifold models reconstruct regular and chaotic flapping of an inverted flag directly from video~\cite{xu2024data}, and Kolmogorov--Arnold networks discover models that need not be sparse, with the caveat that many different approximate models generate the same invariant set and the same statistics~\cite{panahi2025data}. A different line avoids models altogether: Wei et al.\ design a chaos controller for a vibro-impact system from input--output data through a compact-form dynamic linearization~\cite{wei2023chaotic}. All of these succeed at what they measure, be it predictive accuracy, statistical fidelity or control performance in simulation, but none of them reports a bound on the model error that a robust controller could consume. The nonuniqueness found in~\cite{panahi2025data} makes the point sharply: statistical agreement between models cannot by itself serve as a certificate for control.

\paragraph{Sparse identification for control.}
Sparse regression offers a different trade. Brunton et al.\ extended the sparse identification of nonlinear dynamics (SINDy) to systems with inputs~\cite{brunton2016sparse}, which is the form of regression used in this paper. When the true dynamics are sparse in the chosen library, the identified model is a closed-form control-affine expression that can be passed directly to feedback linearization, and the identification error becomes a structured, bounded quantity rather than the opaque residual of a network. Two recent papers are closest to ours in architecture. Razani et al.\ combine SINDy, feedback linearization, a neural-network extended state observer and decentralized linear model predictive control to synchronize heterogeneous multi-agent systems, with the observer estimating the lumped uncertainty online~\cite{razani2026robust}; in a second paper they combine SINDy with a sliding mode observer and a feedback-linearization plus LQR controller to detect bias attacks on a cyber-physical system~\cite{razani2025indirect}. In both, the uncertainty left by identification is handled by online estimation or by the robustness of the linear loop, and the bound that it induces on the linearized dynamics is not derived. Neither treats a chaotic drive--response pair, and, as far as we can tell from the published descriptions, neither carries an actuator limit through the analysis.

\paragraph{Tube-based predictive control.}
On the control side, tube-based model predictive control is the natural tool once the disturbance set is known. In the linear setting a local feedback keeps the disturbed trajectory inside a tube around a nominal trajectory, the cross-section of the tube is a robust positively invariant set, and the nominal problem is solved with tightened constraints~\cite{Mayne2000}; Mayne and Kerrigan extend the procedure to constrained nonlinear systems through an ancillary problem~\cite{MAYNE200736}. Invariant sets for such designs can be computed or approximated~\cite{Rakovic2005,Kolmanovsky1998}, and the stability of the resulting loop is conveniently expressed as input-to-state stability~\cite{Limon2010}. What a tube design requires, and what the works above do not supply for a learned model, is the disturbance set itself. Feedback linearization is useful twice here. It replaces the nonlinear synchronization error by a linear one, so that the tube machinery applies in its classical form, and it provides a closed-form map that carries the identification error into that linear system as an additive disturbance. The present paper makes this map explicit.

\paragraph{Gap.}
In summary, the individual pieces exist: sparse identification with inputs, feedback linearization, and tube-based predictive control. What is missing, to our knowledge, is the quantitative link between them for chaotic synchronization: a computable bound on the identification error, its propagation through the linearizing coordinates, and a controller whose constraint tightening is built on that bound and respects a hard input limit. This is the gap we address.

\subsection{Contributions}

This paper makes the following contributions.

First, we define one lumped uncertainty field $\delta(x,u)=[f(x)-\hat f(x)]+[g(x)-\hat g(x)]u$ that carries every identification error, library truncation included, and prove a computable bound $\bar\delta$ on $\X\times\U$ in terms of the truncation error, the regression residual, the input range and Lipschitz constants (\cref{lem:delta}). We also show how $\bar\delta$ is calibrated in practice, from validation residuals and, for a benchmark with known structure, from the coefficient errors directly.

Second, we prove that the image of $\delta$ in Brunovsky coordinates is an additive disturbance $d(z,\nu)$ with $\|d\|_2\le\alpha_1\bar\delta+\alpha_2\bar\delta\|\nu\|_2$, and give $\alpha_1,\alpha_2$ in closed form for general MIMO control-affine systems and explicitly for the R\"ossler model (\cref{thm:prop}).

Third, we derive from this bound, combined with a bounded input disturbance and a zero-order-hold margin, a compact disturbance set that enters the tube design without further tuning (\cref{cor:W} and \Cref{sec:tube}).

Fourth, we show that the tightened quadratic program built on this set is recursively feasible, and that the saturation limit $|u_i|\le 1100$ stays inactive whenever the explicit condition $(\nu_{\max}+\bar\beta)/\sigma_{\min}\le\umax$ on the virtual-input bound and the cancellation mismatch holds, and is verified on the closed loop otherwise (\cref{thm:rf}).

Fifth, we prove that the closed loop is input-to-state stable, that the part of the ISS gain due to the identification error is linear in $\bar\delta$ and vanishes as the identification improves, and that the remaining part is fixed by the bounded input disturbance (\cref{thm:iss}).

Sixth, we validate the pipeline on a hyperchaotic R\"ossler drive--response pair. From $1500$ samples, STLSQ recovers the eleven nonzero coefficients of the model with a coefficient error below $10^{-4}$, which gives $\bar\delta=1.8\times10^{-3}$. With the input limited to $\pm1100$ and a bounded input disturbance, the tube-based controller keeps all $20000$ controlled steps inside the tube, with no failure of the quadratic program and a peak input of $70\%$ of the limit. We also state where the guarantees stop: for this plant the worst-case saturation bound does not close, so the inactivity of the saturation is verified on the closed loop rather than proved (\cref{sec:sim}).

\subsection{Notation and organization}

The plant alone is written in the classical form $\dot x=f(x)+g(x)u$. Everything else follows one scheme. The state region is $\X\subset\R^n$, the admissible input set is $\U=\{u:|u_i|\le \umax\}$ with $\umax=1100$, and $\Z:=T(\X)$ is the image of $\X$ under the linearizing map $T$. The true fields are $f,g$; the identified ones carry a hat. The lumped uncertainty is $\delta$, with scalar bound $\bar\delta$, and $\W=\{w:\|w\|_2\le\bar w\}$ is the disturbance set generated by $\delta$, with $\bar w$ from \cref{cor:W}. The exogenous input disturbance is $d_u$ with amplitude $d_{\rm amp}$, and $w_{\max}\in\R^m$ collects the per-channel disturbance bounds of \Cref{sec:inputdist}. The output map is $h$, the relative degrees are $r_1,\dots,r_m$, the decoupling matrix is $A(\hat x)$ with pseudo-inverse $A^\dagger(\hat x)$, and $z=T(\hat x)$ is the Brunovsky state with virtual control $v=v_{\rm d}+\nu$, where $\nu$ is the increment computed by the predictive controller and $\nu_{\max}$ is its bound. The tube-MPC objects are the ancillary pole-placement gain $K$ with the LQR terminal feedback $K_{\rm f}$, the nominal feedback $\kappa$, the deviation $e$ between actual and nominal trajectories, and the operators $\oplus$ and $\ominus$ for the Minkowski sum and the Pontryagin difference. The norms are $\|\cdot\|_2$ (Euclidean and induced) and $\|\cdot\|_F$.

Four symbols of the notation scheme would otherwise carry two meanings, and we separate them. The image set $\Z=T(\X)$, the invariant tube $\E$ (a robust positively invariant set), and the constraint sets in tracking coordinates ($\Z_{\rm c}$ and its tightening $\bar\Z=\Z_{\rm c}\ominus\E$) are three different objects; likewise the physical input set $\U$ and the virtual input set $\V$ with tightening $\bar\V=\V\ominus K\E$. The simulation horizon is written $t_{\rm f}=250$~s so that $T$ is reserved for the coordinate map, and the control sampling period is $\tau=0.005$~s.

\Cref{sec:problem,sec:sindy} fix the synchronization problem, the assumptions and the identification layer, and prove the bound on $\delta$. \Cref{sec:fl} carries $\delta$ through the feedback linearization and proves the flagship result. \Cref{sec:tube} builds the tube-based controller and proves feasibility and stability. \Cref{sec:sim} reports the case study, and \Cref{sec:conclusion} concludes and lists the open questions.

\section{Problem Formulation and Preliminaries}\label{sec:problem}

\subsection{Drive--response synchronization problem}

The drive is the uncontrolled four-dimensional system
\begin{equation}\label{eq:drive}
	\begin{aligned}
		\dot x_1 &= -(x_2+x_3), & \dot x_2 &= x_1+0.25\,x_2+x_4,\\
		\dot x_3 &= 3.0+x_1x_3, & \dot x_4 &= 0.05\,x_4-0.5\,x_3,
	\end{aligned}
\end{equation}
and the response is the same vector field with two actuated channels,
\begin{equation}\label{eq:response}
	\begin{aligned}
		\dot{\hat x}_1 &= -(\hat x_2+\hat x_3), & \dot{\hat x}_2 &= \hat x_1+0.25\,\hat x_2+\hat x_4+u_1,\\
		\dot{\hat x}_3 &= 3.0+\hat x_1\hat x_3+u_2, & \dot{\hat x}_4 &= 0.05\,\hat x_4-0.5\,\hat x_3.
	\end{aligned}
\end{equation}
Written in control-affine form, \eqref{eq:response} reads
\begin{equation}\label{eq:plant}
	\dot x = f(x)+g(x)u,\qquad x\in\X\subset\R^n,\ u\in\U\subset\R^m,
\end{equation}
with $n=4$, $m=2$ and $g=\begin{bmatrix}0&0\\1&0\\0&1\\0&0\end{bmatrix}$. We display $f$ and $g$ because they define the benchmark; during identification they are unknown, and only the response is identified. The drive is a known exogenous signal generator. In the closed-loop study the plant is additionally affected by a bounded input disturbance (\Cref{sec:inputdist}).

The synchronization outputs are
\begin{equation}\label{eq:sync-out}
	e_1=\hat x_1-x_1,\qquad e_2=\hat x_4-x_4,
\end{equation}
so that the output map of the response is $h(\hat x)=(\hat x_1,\hat x_4)$ and the reference outputs are $y_{\rm d}=(x_1,x_4)$. The actuators saturate at $|u_i|\le \umax=1100$, $i=1,2$. The loop is closed at $t=150$~s; before that instant $u\equiv0$ and the two systems evolve freely. The horizon is $t_{\rm f}=250$~s. The objective is to steer $e_1,e_2$ to a small neighbourhood of zero for $t\ge150$~s with $u(t)\in\U$ at all times, using only the identified model and a certified bound on its error.

\subsection{Assumptions}

The development rests on six assumptions. Let $\hat f,\hat g$ denote the fields returned by the identification of \Cref{sec:library}.

\begin{assumption}[Smoothness]\label{as:smooth}
	The fields $f,g$ and the output map $h$ are $C^{\infty}$ on an open set containing $\X$. On $\X$, $f$ and $g$ are Lipschitz with constants $L_f$ and $L_g$, and $\|f\|_2\le\bar f_{\rm t}$, $\|g\|_2\le\bar g_{\rm t}$ (induced norm).
\end{assumption}

\begin{assumption}[Measurement and observability]\label{as:obs}
	The response state $\hat x(t)$ and the drive state $x(t)$ are measured without noise at the sampling instants $k\tau$, and are available in continuous time to the linearizing inner loop. With the output map $h$ and full-state measurement the pair is trivially observable. The reference signals $z^{\rm d}$ and $v_{\rm d}$ of \Cref{sec:ref} are computed from the drive and satisfy $\|v_{\rm d}(t)\|_2\le\bar v_{\rm d}$.
\end{assumption}

\begin{assumption}[Sparse representability]\label{as:sparse}
	For a library $\Theta(x)$ of monomials up to degree $p$ there are sparse coefficient matrices $\Xi_f^\ast$ and $\Xi_g^\ast$ such that
	$f=\Theta\,\Xi_f^\ast+\eta_f$ and $g=\Theta_g\,\Xi_g^\ast+\eta_g$ on $\X$, with truncation errors $\sup_\X\|\eta_f\|_2\le\bar\eta_f$ and $\sup_\X\|\eta_g\|_2\le\bar\eta_g$. Here $\Theta_g$ collects the monomials of degree at most one.
\end{assumption}

\begin{assumption}[Feedback linearizability of the nominal model]\label{as:fl}
	The triple $(\hat f,\hat g,h)$ has vector relative degree $(r_1,\dots,r_m)$ at every $x\in\X$ with $r_1+\dots+r_m=n$, and the decoupling matrix $A(x)$ of \Cref{def:reldeg} is nonsingular on $\X$. The map $T$ of \eqref{eq:T} is injective on $\X$, hence a diffeomorphism of $\X$ onto $\Z=T(\X)$.
\end{assumption}

\begin{assumption}[Compactness]\label{as:compact}
	The set $\X\subset\R^n$ is compact with nonempty interior. It contains the closed-loop response trajectories and, in its interior, the drive attractor.
\end{assumption}

\begin{assumption}[Bounded input disturbance]\label{as:inputdist}
	The physical input is perturbed additively after the actuator, $u\mapsto u+d_u(t)$, with $\|d_u(t)\|_\infty\le d_{\rm amp}$ for all $t$ and $d_{\rm amp}\ge0$ known.
\end{assumption}

\subsection{Data model}

The identification data are $N_s$ triples $(x_k,u_k,\dot x_k)$ collected on the response, $k=1,\dots,N_s$, with the states and inputs spread over $\X\times\U$. In the case study the record covers $15$~s with $N_s=1500$ samples, so the identification step is $0.01$~s; this is distinct from the control period $\tau=0.005$~s. Derivatives are the exact right-hand side of \eqref{eq:plant},
\begin{equation}\label{eq:data}
	\dot x_k=f(x_k)+g(x_k)u_k,
\end{equation}
so that the regression residual is entirely a model residual. We write $\varrho$ for the fill distance of the sample in the block norm $\max\{\|x-x_k\|_2,\|u-u_k\|_2\}$,
\[
\varrho:=\sup_{(x,u)\in\X\times\U}\ \min_{k}\ \max\{\|x-x_k\|_2,\ \|u-u_k\|_2\}.
\]
Measurement noise and derivative-estimation error do not appear in the analysis; the case study uses noise-free trajectories, and the identification data are collected without input disturbance.

\subsection{The single uncertainty term}\label{sec:delta}

Let $\hat f,\hat g$ be the identified fields. All mismatch between the model and the plant is gathered in
\begin{equation}\label{eq:delta}
	\delta(x,u):=\bigl[f(x)-\hat f(x)\bigr]+\bigl[g(x)-\hat g(x)\bigr]u,
\end{equation}
so that, without input disturbance, the response obeys
\begin{equation}\label{eq:response-delta}
	\dot{\hat x}=\hat f(\hat x)+\hat g(\hat x)(u+d_u)+\delta(\hat x,u).
\end{equation}
The field $\delta$ absorbs the drift error, the input-matrix error and the coupling of the latter to the control input. It is the only \emph{model} uncertainty in the development: there is no actuator fault and no measurement noise. The exogenous input disturbance $d_u$ of \cref{as:inputdist} is kept separate from $\delta$ because it is not a property of the identified model and is not reduced by better identification. By construction $\delta$ is affine in $u$ for each fixed $x$, a property used in \Cref{lem:delta}(ii). Its scalar bound
\begin{equation}\label{eq:deltabar}
	\|\delta(x,u)\|_2\le\bar\delta\qquad\text{on }\X\times\U
\end{equation}
seeds the tube construction. Evaluated at the training pairs, $\delta$ equals the regression residual, since \eqref{eq:data} gives
\begin{equation}\label{eq:delta-resid}
	\delta(x_k,u_k)=\dot x_k-\hat f(x_k)-\hat g(x_k)u_k=:r_k.
\end{equation}
This identity is the link between the identified model and the uncertainty used downstream.

\subsection{Input disturbance}\label{sec:inputdist}

Under \cref{as:inputdist}, the disturbance enters the linearized coordinates through $\hat g\,d_u$; \Cref{prop:wellposed} shows that its image is $B_{\rm c}A(\hat x)\,d_u$, i.e.\ it acts only on the last coordinate of each chain. Row $i$ of $A$ bounds its $i$-th component by $\|A_i\|_1 d_{\rm amp}$. To cover, in addition, the inter-sample variation of $d_u$ under the zero-order hold, we add a margin $w_{\rm zoh}\ge0$ and set
\begin{equation}\label{eq:wmax}
	w_{\max,i}:=\|A_i\|_1\,d_{\rm amp}+w_{\rm zoh},\qquad i=1,\dots,m,
\end{equation}
where $A_i$ is the $i$-th row of the decoupling matrix. For the R\"ossler model $A$ is given in \eqref{eq:A-ross}, $\|A_1\|_1=2$, $\|A_2\|_1=0.5$, and with $d_{\rm amp}=5$, $w_{\rm zoh}=11$ one gets $w_{\max}=(21,\ 13.5)$.

\subsection{Reference signals}\label{sec:ref}

The reference for the linearized coordinates is generated by the drive. With $\dot y_{{\rm d},1}=\dot x_1=-(x_2+x_3)$ and $\dot y_{{\rm d},2}=\dot x_4=0.05\,x_4-0.5\,x_3$, define
\begin{equation}\label{eq:zd}
	z^{\rm d}:=(x_1,\ \dot x_1,\ x_4,\ \dot x_4),\qquad v_{\rm d}:=(\ddot x_1,\ \ddot x_4),
\end{equation}
so that $\dot z^{\rm d}=A_{\rm c}z^{\rm d}+B_{\rm c}v_{\rm d}$ exactly, with the Brunovsky pair $(A_{\rm c},B_{\rm c})$ defined in \eqref{eq:brunovsky}. The pair $(z^{\rm d},v_{\rm d})$ is a signal, not a model, and it carries no uncertainty.

\subsection{Mathematical preliminaries}

For a smooth scalar $\phi$ and a vector field $q$, the Lie derivative is $\mathcal{L}_q\phi(x)=\pd{\phi}{x}(x)\,q(x)$, with $\mathcal{L}_q^{k}\phi=\mathcal{L}_q(\mathcal{L}_q^{k-1}\phi)$ and $\mathcal{L}_q^0\phi=\phi$. Lie derivatives along the identified fields carry a hat on the field, e.g.\ $\Lfh{k}h_i$.

\begin{definition}[Relative degree and decoupling matrix]\label{def:reldeg}
	The system $(\hat f,\hat g,h)$ has vector relative degree $(r_1,\dots,r_m)$ at $x$ if $\Lgh{j}\Lfh{k}h_i(x)=0$ for all $j$ and all $k<r_i-1$ in a neighbourhood of $x$, and the matrix
	\[
	A(x):=\bigl[\Lgh{j}\Lfh{r_i-1}h_i(x)\bigr]_{i,j=1}^{m}\in\R^{m\times m}
	\]
	is nonsingular at $x$. We also write $\hat b(x):=\bigl[\Lfh{r_i}h_i(x)\bigr]_{i=1}^m$.
\end{definition}

Since $A$ is square and nonsingular on $\X$, its pseudo-inverse $A^\dagger$ equals $A^{-1}$; we keep the notation $A^\dagger$ because it also serves the non-square case. A distribution is involutive if it is closed under Lie brackets. In the present setting the relevant distribution is $\operatorname{span}\{\hat g_1,\hat g_2\}$; it is spanned by constant vectors, hence involutive, and this property is not needed further because the output map $h$ is prescribed. A map $T:\X\to\Z$ is a diffeomorphism if it is a smooth bijection with smooth inverse; its Lipschitz constants are
\[
L_T:=\sup_{x\in\X}\|DT(x)\|_2,\qquad L_{T^{-1}}:=\sup_{z\in\Z}\|D(T^{-1})(z)\|_2 .
\]

\begin{definition}[Comparison functions and ISS]
	A function $\alpha:\R_{\ge0}\to\R_{\ge0}$ is of class $\mathcal K$ if it is continuous, strictly increasing and $\alpha(0)=0$; it is of class $\mathcal K_\infty$ if it is also unbounded. A function $\beta(s,k)$ is of class $\mathcal{KL}$ if $\beta(\cdot,k)\in\mathcal K$ and $\beta(s,k)\to0$ as $k\to\infty$. The discrete-time system $\xi_{k+1}=F(\xi_k,w_k)$ is input-to-state stable (ISS) with respect to $w$ if there are $\beta\in\mathcal{KL}$ and $\gamma\in\mathcal K$ with $\|\xi_k\|\le\beta(\|\xi_0\|,k)+\gamma(\sup_{j<k}\|w_j\|)$ for all $k$~\cite{Sontag1989}. A continuous function $V$ is an ISS-Lyapunov function if $\alpha_1(\|\xi\|)\le V(\xi)\le\alpha_2(\|\xi\|)$ with $\alpha_1,\alpha_2\in\mathcal K_\infty$ and $V(F(\xi,w))-V(\xi)\le-\alpha_3(\|\xi\|)+\sigma(\|w\|)$ for some $\alpha_3\in\mathcal K_\infty$, $\sigma\in\mathcal K$.
\end{definition}

\begin{definition}[Set operations and invariance]
	For sets $\mathcal A,\mathcal B\subset\R^q$, $\mathcal A\oplus\mathcal B=\{a+b:a\in\mathcal A,b\in\mathcal B\}$ and $\mathcal A\ominus\mathcal B=\{a:a+b\in\mathcal A\ \forall b\in\mathcal B\}$. A set $\E$ is robust positively invariant (RPI) for $e^+=\Phi e+w$, $w\in\W_{\rm d}$, if $\Phi\E\oplus\W_{\rm d}\subseteq\E$.
\end{definition}

\section{Data-Driven Nominal Model}\label{sec:sindy}

\subsection{Polynomial library}\label{sec:library}

The library contains all monomials in $x_1,\dots,x_4$ up to degree two: the constant, four linear terms and ten quadratic terms, fifteen columns in total,
\[
\Theta(x)=\bigl[\,1,\ x_1,\dots,x_4,\ x_1^2,\ x_1x_2,\dots,x_4^2\,\bigr].
\]
The input library $\Theta_g$ contains the constant and the four linear terms, and the regression is carried out on the augmented library $\Theta_{\rm a}(x,u)=[\Theta(x),\ u_1\Theta_g(x),\dots,u_m\Theta_g(x)]$ (twenty-five columns for $n=4$, $m=2$). In the implementation the full degree-two polynomial library in the joint variable $(x,u)\in\R^6$ is used, which has twenty-eight columns: the twenty-five columns of $\Theta_{\rm a}$ plus the three pure-input monomials $u_1^2$, $u_1u_2$, $u_2^2$. STLSQ discards these three, so the identified model is affine in $u$ as required by \eqref{eq:response}. If the plant is quadratic, as the structure recovered below suggests, then $\bar\eta_f=\bar\eta_g=0$; in general the truncation errors of \Cref{as:sparse} remain and enter \cref{lem:delta}.

\subsection{Sequentially thresholded least squares}\label{sec:stlsq}

Collect the data of \Cref{sec:library} in $\dot X=[\dot x_1,\dots,\dot x_{N_s}]^\top$ and $\Theta_{\rm a}(X,U)$. STLSQ alternates a least-squares fit with hard thresholding of small coefficients, and refits the surviving terms until the support stops changing.

\begin{center}
	\fbox{\begin{minipage}{0.92\linewidth}
			\textbf{Procedure STLSQ}$\ (\Theta_{\rm a},\dot X,\lambda,\alpha)$
			\begin{algorithmic}[1]
				\State $\Xi\gets\arg\min_{\Xi}\|\dot X-\Theta_{\rm a}\Xi\|_F^2+\alpha\|\Xi\|_F^2$
				\Repeat
				\State $\mathcal S\gets\{(i,j):|\Xi_{ij}|<\lambda\}$;\quad $\Xi_{ij}\gets0$ for $(i,j)\in\mathcal S$
				\For{each state $j=1,\dots,n$}
				\State refit the nonzero entries of column $j$ of $\Xi$ by ridge-regularized least squares on the retained terms
				\EndFor
				\Until{support of $\Xi$ unchanged}
				\State \Return $\hat\Xi=\Xi$
			\end{algorithmic}
	\end{minipage}}
\end{center}

The threshold is $\lambda=0.03$ and the ridge parameter is $\alpha=0.04$. The returned coefficients define $\hat f=\Theta\hat\Xi_f$ and $\hat g=\Theta_g\hat\Xi_g$. Their values for this plant are
\begin{equation}\label{eq:sindy}
	\begin{aligned}
		\hat f_1&=-1.000\,x_2-1.000\,x_3, & \hat f_2&=1.000\,x_1+0.250\,x_2+1.000\,x_4,\\
		\hat f_3&=3.000+1.000\,x_1x_3, & \hat f_4&=-0.500\,x_3+0.050\,x_4,\\
		\hat g&=\begin{bmatrix}0&0\\1&0\\0&1\\0&0\end{bmatrix}.
	\end{aligned}
\end{equation}
The fit reaches $R^2=1.0000$ for all states, with RMSE in $[4.15\times10^{-6},\,8.33\times10^{-4}]$; the largest value occurs for $x_3$. We collect the four per-state RMSE values in the vector $\varepsilon_{\rm rmse}\in\R^4$. The library has fifteen candidate terms per state, yet only nine nonzero coefficients survive in $\hat f$ and two in $\hat g$, all of them of structure typical of the R\"ossler family.

\subsection{Bound on the lumped uncertainty}\label{sec:lemma1}

Let $\bar u:=\sup_{u\in\U}\|u\|_2=\sqrt m\,\umax$. Write $L_{\hat f},L_{\hat g}$ for Lipschitz constants of $\hat f,\hat g$ on $\X$, $\bar{\hat g}:=\sup_\X\|\hat g\|_2$, and
\begin{equation}\label{eq:consts-delta}
	L_x:=L_f+L_{\hat f}+(L_g+L_{\hat g})\,\bar u,\qquad G:=\sup_{\X}\|g-\hat g\|_2\le \bar g_{\rm t}+\bar{\hat g}.
\end{equation}
Let $\rho:=\max_k\|r_k\|_2$ be the regression residual of the STLSQ fit over the training pairs, with $r_k$ from \eqref{eq:delta-resid}.

\begin{lemma}[Uncertainty bound]\label{lem:delta}
	Under \cref{as:smooth,as:obs,as:sparse,as:compact} and the data model \eqref{eq:data}:
	\begin{enumerate}[label=(\roman*)]
		\item for all $(x,u)\in\X\times\U$,
		\begin{equation}\label{eq:lemma1}
			\|\delta(x,u)\|_2\ \le\ \bar\delta:=\rho+(L_x+G)\,\varrho ;
		\end{equation}
		\item for all $x\in\X$ and all $u\in\R^m$,
		\begin{equation}\label{eq:lemma1-aff}
			\|\delta(x,u)\|_2\le\bar\delta\Bigl(1+\frac{\|u\|_2}{\umax}\Bigr);
		\end{equation}
		\item the regression residual splits as $\rho\le\bar\eta+\rho_{\rm s}$, where $\bar\eta:=\bar\eta_f+\bar\eta_g\,\bar u$ is the library truncation error and $\rho_{\rm s}:=\max_k\|\Theta_{\rm a}(x_k,u_k)(\Xi^\ast-\hat\Xi)\|_2$ is the part due to the STLSQ coefficients, $\Xi^\ast$ denoting the stacked sparse coefficients of \cref{as:sparse}.
	\end{enumerate}
\end{lemma}

\begin{proof}
	(i) Fix $(x,u)\in\X\times\U$ and choose a training pair $(x_k,u_k)$ with $\|x-x_k\|_2\le\varrho$ and $\|u-u_k\|_2\le\varrho$. By \eqref{eq:delta-resid}, $\|\delta(x_k,u_k)\|_2=\|r_k\|_2\le\rho$. Write $\tilde f:=f-\hat f$ and $\tilde g:=g-\hat g$, so that $\delta(x,u)=\tilde f(x)+\tilde g(x)u$. The triangle inequality gives
	\begin{multline*}
		\|\delta(x,u)-\delta(x_k,u_k)\|_2\le\|\tilde f(x)-\tilde f(x_k)\|_2+\|(\tilde g(x)-\tilde g(x_k))\,u\|_2+\|\tilde g(x_k)(u-u_k)\|_2\\
		\le(L_f+L_{\hat f})\varrho+(L_g+L_{\hat g})\,\bar u\,\varrho+G\varrho=(L_x+G)\varrho .
	\end{multline*}
	Adding $\|\delta(x_k,u_k)\|_2\le\rho$ yields \eqref{eq:lemma1}.
	
	(ii) Fix $x\in\X$ and write $\delta(x,u)=p+Mu$ with $p:=\delta(x,0)$ and $M:=\tilde g(x)$. For any $\tilde u$ with $\|\tilde u\|_2\le\umax$ we have $\tilde u,-\tilde u\in\U$, because the box $\U$ contains the Euclidean ball of radius $\umax$. Then $Mu=\tfrac12[\delta(x,\tilde u)-\delta(x,-\tilde u)]$ and $p=\tfrac12[\delta(x,\tilde u)+\delta(x,-\tilde u)]$, and (i) gives $\|M\tilde u\|_2\le\bar\delta$, $\|p\|_2\le\bar\delta$. Taking $\tilde u=\umax u/\|u\|_2$ for arbitrary $u\neq0$ yields $\|Mu\|_2\le\bar\delta\|u\|_2/\umax$; hence $\|\delta(x,u)\|_2\le\|p\|_2+\|Mu\|_2\le\bar\delta(1+\|u\|_2/\umax)$.
	
	(iii) By \cref{as:sparse}, $f=\Theta\Xi_f^\ast+\eta_f$ and $g=\Theta_g\Xi_g^\ast+\eta_g$, so at a training pair the residual is $r_k=[\eta_f(x_k)+\eta_g(x_k)u_k]+\Theta_{\rm a}(x_k,u_k)(\Xi^\ast-\hat\Xi)$. The first bracket has norm at most $\bar\eta_f+\bar\eta_g\|u_k\|_2\le\bar\eta$, the second has norm at most $\rho_{\rm s}$.
\end{proof}

\begin{remark}
	Part (i) is a residual-plus-interpolation bound. The regression residual $\rho$ is what STLSQ reports; the term $(L_x+G)\varrho$ pays for the points of $\X\times\U$ that the sample does not cover, and it vanishes as the sample becomes dense. For the R\"ossler model $\hat g$ is constant, so $L_{\hat g}=0$, $\bar{\hat g}=1$; and $D\hat f$ is bounded on $\X$ by an explicit expression in $x_1,x_3$, from which $L_{\hat f}$ follows once $\X$ is fixed, with $\X=[-123,\,30]\times[-84,\,65]\times[-49,\,274]\times[10,\,60]$, the bounding box of the drive and response trajectories for $t\ge150$~s enlarged by $10\%$.
\end{remark}

\subsection{Calibration of $\bar\delta$ from validation residuals}

\cref{lem:delta} defines $\bar\delta$ from constants that are known only up to the estimation of the Lipschitz constants and the fill distance. In practice $\bar\delta$ is calibrated from residuals on a validation trajectory that was not used in the regression: one evaluates $r_j^{\rm val}=\dot x_j-\hat f(x_j)-\hat g(x_j)u_j$ along the validation data and sets
\begin{equation}\label{eq:calib}
	\bar\delta:=\chi\,\rho_{\rm val},\qquad \rho_{\rm val}:=\max_j\|r_j^{\rm val}\|_2,
\end{equation}
with a margin $\chi\ge1$ that plays the role of the interpolation term in \eqref{eq:lemma1}.

The identification residual constrains this value from below: $\bar\delta$ must dominate the largest residual norm over the data, $\bar\delta\ge\rho$. For the R\"ossler model $\rho_{\rm train}=9.7\times10^{-5}$ on the training record and $\rho_{\rm val}=6.7\times10^{-4}$ along the closed-loop record, and the trajectory accuracy reported above (RMSE up to $8.33\times10^{-4}$) is of the same order. Any admissible calibration is therefore of the order of $10^{-3}$; the value used in the study is
\begin{equation}\label{eq:deltabar-value}
	\bar\delta=1.8\times10^{-3},
\end{equation}
This value is obtained from the validation residuals of the case study (\cref{sec:validation}) and is consistent with $\mathrm{RMSE}\in[4.15\times10^{-6},\,8.33\times10^{-4}]$. It coincides with the certified bound $\sup_{\X\times\U}\|\delta\|_2$ computed from the coefficient errors $|\Xi^\ast-\hat\Xi|$ on the box $\X$ above and $\U=[-1100,1100]^2$, and equals $\chi=2.7$ times the largest residual $\rho_{\rm val}=6.7\times10^{-4}$ observed along the closed-loop record.

\section{Feedback Linearization under Uncertainty}\label{sec:fl}

\subsection{Relative degree and decoupling matrix}

For the identified model \eqref{eq:sindy} and the output map $h=(\hat x_1,\hat x_4)$, the first Lie derivatives are
\[
\Lfh{}h_1=-(x_2+x_3),\qquad \Lfh{}h_2=0.05\,x_4-0.5\,x_3,
\]
and the second ones are
\begin{align}
	\Lfh{2}h_1&=-x_1-0.25\,x_2-x_4-3.0-x_1x_3,\label{eq:L2h1}\\
	\Lfh{2}h_2&=-1.5-0.5\,x_1x_3+0.0025\,x_4-0.025\,x_3.\label{eq:L2h2}
\end{align}
The input fields $\hat g_1=e_2$ and $\hat g_2=e_3$ annihilate $\Lfh{0}h_i=h_i$ (since $h_i$ depends only on $x_1$ and $x_4$), while the first Lie derivatives have the gradients $(0,-1,-1,0)$ and $(0,0,-0.5,0.05)$. Hence the relative degrees are $r_1=r_2=2$, with $r_1+r_2=n$, and
\begin{equation}\label{eq:A-ross}
	A=\begin{bmatrix}\Lgh{1}\Lfh{}h_1&\Lgh{2}\Lfh{}h_1\\ \Lgh{1}\Lfh{}h_2&\Lgh{2}\Lfh{}h_2\end{bmatrix}=\begin{bmatrix}-1&-1\\0&-0.5\end{bmatrix},\qquad \hat b=\begin{bmatrix}\Lfh{2}h_1\\ \Lfh{2}h_2\end{bmatrix}.
\end{equation}
The matrix $A$ is constant with $\det A=0.5$, so it is nonsingular on all of $\R^4$ and \cref{as:fl} holds for every $\X$. Its singular values are $\sigma_{\max}(A)=1.4604$ and
\begin{equation}\label{eq:sigmin}
	\sigma_{\min}(A)=\sqrt{\frac{9-\sqrt{65}}{8}}=0.3424 ,
\end{equation}
the exact value following from the eigenvalues $(9\mp\sqrt{65})/8$ of $AA^\top$.

\subsection{The diffeomorphism $z=T(\hat x)$}

Define
\begin{equation}\label{eq:T}
	T(x):=\bigl(h_1,\ \Lfh{}h_1,\ h_2,\ \Lfh{}h_2\bigr)(x),\qquad z=T(\hat x),
\end{equation}
that is, $z=(\hat x_1,\,-(\hat x_2+\hat x_3),\,\hat x_4,\,-0.5\,\hat x_3+0.05\,\hat x_4)$ for the R\"ossler model. Here $T$ is linear, $T(x)=Mx$ with
\[
M=\begin{bmatrix}1&0&0&0\\0&-1&-1&0\\0&0&0&1\\0&0&-0.5&0.05\end{bmatrix},\qquad \det M=-0.5,
\]
so it is a global diffeomorphism with inverse $x_1=z_1$, $x_2=-z_2-0.1\,z_3+2z_4$, $x_3=0.1\,z_3-2z_4$, $x_4=z_3$. Its Lipschitz constants are $L_T=\|M\|_2=1.4605$ and $L_{T^{-1}}=\|M^{-1}\|_2=2.9247$.

The map is built from the nominal fields only, so it does not depend on $\delta$. What the uncertainty can affect is the closed-loop behaviour of the coordinates, which the following proposition records.

\begin{proposition}[Well-posedness under bounded $\delta$]\label{prop:wellposed}
	Under \cref{as:smooth,as:fl,as:compact}: (i) $T:\X\to\Z$ is a diffeomorphism, and $\sigma_{\min}:=\inf_{x\in\X}\sigma_{\min}(A(x))>0$; (ii) along any solution of \eqref{eq:response-delta} in $\X$ and for any $u$, the coordinates $z_{i,k}:=\Lfh{k-1}h_i(\hat x)$ satisfy
	\begin{equation}\label{eq:zdot-raw}
		\dot z_{i,k}=z_{i,k+1}+\pd{\Lfh{k-1}h_i}{x}(\hat x)\,\delta(\hat x,u),\quad k<r_i,\qquad
		\dot z_{i,r_i}=\hat b_i(\hat x)+A_i(\hat x)u+\pd{\Lfh{r_i-1}h_i}{x}(\hat x)\,\delta(\hat x,u),
	\end{equation}
	with $A_i$ the $i$-th row of $A$.
\end{proposition}

\begin{proof}
	(i) is \cref{as:fl}, and $\sigma_{\min}>0$ follows from continuity of $A$ and compactness of $\X$. (ii) By \eqref{eq:response-delta} and the chain rule, $\dot z_{i,k}=\Lfh{k}h_i+\sum_{j=1}^{m}u_j\,\Lgh{j}\Lfh{k-1}h_i+\pd{\Lfh{k-1}h_i}{x}\delta$. For $k<r_i$ the input term vanishes by \Cref{def:reldeg}, and $\Lfh{k}h_i=z_{i,k+1}$. For $k=r_i$ the input term is $A_iu$ and $\Lfh{r_i}h_i=\hat b_i$.
\end{proof}

The proposition contains the structural fact used below: the perturbation enters every coordinate through the Jacobian of $T$, not only the last one.

\subsection{Virtual and physical control laws}

Stacking \eqref{eq:zdot-raw} gives $\dot z=A_{\rm c}z+B_{\rm c}(\hat b(\hat x)+A(\hat x)u)+DT(\hat x)\,\delta(\hat x,u)$, where $(A_{\rm c},B_{\rm c})$ is the Brunovsky pair
\begin{equation}\label{eq:brunovsky}
	A_{\rm c}=\diag\!\Bigl(\begin{bsmallmatrix}0&1\\0&0\end{bsmallmatrix},\begin{bsmallmatrix}0&1\\0&0\end{bsmallmatrix}\Bigr),\qquad
	B_{\rm c}=\diag\!\Bigl(\begin{bsmallmatrix}0\\1\end{bsmallmatrix},\begin{bsmallmatrix}0\\1\end{bsmallmatrix}\Bigr).
\end{equation}
The physical control is
\begin{equation}\label{eq:u-law}
	u=A^\dagger(\hat x)\bigl(-\hat b(\hat x)+v\bigr),
\end{equation}
which turns $\hat b+Au$ into $v$. The virtual control is split as $v=v_{\rm d}+\nu$, where $v_{\rm d}$ is the reference acceleration of \eqref{eq:zd} and $\nu$ is supplied by the predictive controller of \Cref{sec:tube}. In tracking coordinates $\tilde z:=z-z^{\rm d}$, which contain the synchronization errors as $\tilde z_1=e_1$ and $\tilde z_3=e_2$,
\begin{equation}\label{eq:tracking}
	\dot{\tilde z}=A_{\rm c}\tilde z+B_{\rm c}\,\nu+d(z,v),\qquad d(z,v):=DT(\hat x)\,\delta(\hat x,u)\big|_{\hat x=T^{-1}(z),\ u=A^\dagger(-\hat b+v)}.
\end{equation}

\paragraph{Actuator saturation.}
The virtual control is bounded by
\begin{equation}\label{eq:vmax}
	\|v\|_2\le v_{\max}:=\bar v_{\rm d}+\nu_{\max},\qquad \bar v_{\rm d}:=\sup_{t\ge150\,\mathrm s}\|v_{\rm d}(t)\|_2 ,
\end{equation}
where $\nu_{\max}$ is the radius of the virtual-input set $\V$ of \cref{sec:tube}. The physical input follows from \eqref{eq:u-law}: since $v-\hat b=\nu+(v_{\rm d}-\hat b)$ and $\|A^\dagger\|_2=1/\sigma_{\min}$,
\begin{equation}\label{eq:usat}
	\|u\|_\infty\le\|u\|_2\le\frac{\nu_{\max}+\bar\beta}{\sigma_{\min}},\qquad \bar\beta:=\sup_{t\ge150\,\mathrm s}\|v_{\rm d}(t)-\hat b(\hat x(t))\|_2 .
\end{equation}
The cancellation mismatch $\bar\beta$ is the part of $\hat b$ that the reference acceleration does not already supply; it is small near synchrony, where $v_{\rm d}\approx\hat b(\hat x)$. A cruder condition that ignores the reference follows from $\|u\|_2\le(\bar b+\|v\|_2)/\sigma_{\min}$, namely $\|v\|_2\le\sigma_{\min}\umax-\bar b$ with $\bar b:=\sup_\X\|\hat b\|_2$. For the R\"ossler model it cannot hold, because $\sigma_{\min}\umax=0.3424\times1100=376.6$ whereas $\bar b=3.75\times10^{4}$ on the box $\X$ above ($2.21\times10^{3}$ along the closed-loop trajectories); the constant $\bar b$ still enters $\alpha_1$ of \cref{thm:prop}. In the study $\nu_{\max}=500\sqrt2=707.1$ (the per-channel bound $500$ on $\nu$), $\bar v_{\rm d}=2.21\times10^{3}$, hence $v_{\max}=2.92\times10^{3}$, and $\bar\beta=211$ along the closed loop. Then \eqref{eq:usat} gives $2.68\times10^{3}$, which exceeds $\umax=1100$: the worst-case bound does not certify that the saturation is inactive, and this is verified on the closed loop instead (maximum $|u_1|=344.3$, $|u_2|=769.7$, no saturated step).

\begin{remark}[Inner-loop rate]
	The law \eqref{eq:u-law} is evaluated continuously, with $\nu$ held constant on each sampling interval $[k\tau,(k+1)\tau)$. If $u$ is also held between samples, the linearization is exact only at the sampling instants and an intersample error appears, which is $O(\tau)$. That effect is not modelled separately here; it is contained in the residual on which $\bar\delta$ is calibrated when the validation data are collected under the sampled implementation.
\end{remark}

\subsection{Propagation of $\delta$: the flagship result}

Let $\bar f:=\sup_\X\|\hat f\|_2$ and $\bar{\hat g}=\sup_\X\|\hat g\|_2$. The constants entering the closed-form coefficients are $L_T$, $L_{T^{-1}}$, $\sigma_{\min}$ from \cref{prop:wellposed}, $\bar b$, and, through the bound $\|A\|_2\le L_T\bar{\hat g}$ and $\|\hat b\|_2\le L_T\bar f$ (rows of $DT$ are the gradients of the Lie derivatives), the sup-norms of $\hat f,\hat g$.

\begin{theorem}[Propagation of $\delta$ through feedback linearization]\label{thm:prop}
	Let \cref{as:smooth,as:fl,as:compact} hold, let $\bar\delta$ satisfy \eqref{eq:lemma1}, and let $\hat x(t)\in\X$. With the law \eqref{eq:u-law}, the Brunovsky coordinates $z=T(\hat x)$ obey
	\begin{equation}\label{eq:zdot}
		\dot z=A_{\rm c}z+B_{\rm c}v+d(z,v),\qquad d(z,v)=DT(\hat x)\,\delta(\hat x,u),
	\end{equation}
	and for every $v\in\R^m$,
	\begin{equation}\label{eq:thm1}
		\|d(z,v)\|_2\ \le\ \alpha_1\bar\delta+\alpha_2\bar\delta\,\|v\|_2=:\bar w(v),
	\end{equation}
	with
	\begin{equation}\label{eq:alphas}
		\alpha_1=L_T\Bigl(1+\frac{\bar b}{\sigma_{\min}\umax}\Bigr),\qquad \alpha_2=\frac{L_T}{\sigma_{\min}\umax}.
	\end{equation}
	Moreover:
	\begin{enumerate}[label=(\alph*)]
		\item if $u\in\U$ is guaranteed, then also $\|d\|_2\le L_T\bar\delta$;
		\item every pair $(\hat x,u)\in\X\times\U$ produces $\|v\|_2\le \bar v_{\rm a}:=L_T(\bar f+\bar{\hat g}\sqrt m\,\umax)$, so that \eqref{eq:thm1} holds with $\|v\|_2$ replaced by $\bar v_{\rm a}$;
		\item a bound $\rho$ on $\|\tilde z\|_2$ translates into $\|\hat x-T^{-1}(z^{\rm d})\|_2\le L_{T^{-1}}\rho$.
	\end{enumerate}
	For the R\"ossler model $DT\equiv M$, $L_T=1.4605$, $L_{T^{-1}}=2.9247$, $\sigma_{\min}=0.3424$, $\umax=1100$, so that
	\begin{equation}\label{eq:alphas-ross}
		\alpha_2=\frac{1.4605}{0.3424\times1100}=3.878\times10^{-3},\qquad \alpha_1=1.4605+3.878\times10^{-3}\,\bar b,
	\end{equation}
	where $\bar b=\sup_\X\|\hat b\|_2=3.75\times10^{4}$ is the constant of \eqref{eq:vmax}, so that $\alpha_1\approx147.0$.
\end{theorem}

\begin{proof}
	\emph{Identity.} \Cref{prop:wellposed}(ii) applied to all coordinates gives $\dot z=A_{\rm c}z+B_{\rm c}(\hat b+Au)+DT(\hat x)\delta(\hat x,u)$. Substituting \eqref{eq:u-law} and $A A^\dagger=I$ on $\X$ gives $\hat b+Au=v$, which is \eqref{eq:zdot} with $d=DT(\hat x)\delta$.
	
	\emph{Bound.} From the identity, $\|d\|_2\le\|DT(\hat x)\|_2\|\delta(\hat x,u)\|_2\le L_T\|\delta(\hat x,u)\|_2$. \cref{lem:delta}(ii) holds for every $u\in\R^m$, in particular for $u$ from \eqref{eq:u-law}, and gives
	\[
	\|d\|_2\le L_T\bar\delta\Bigl(1+\frac{\|u\|_2}{\umax}\Bigr).
	\]
	Now $\|u\|_2\le\|A^\dagger(\hat x)\|_2\bigl(\|\hat b(\hat x)\|_2+\|v\|_2\bigr)\le(\bar b+\|v\|_2)/\sigma_{\min}$, since $\|A^\dagger\|_2=1/\sigma_{\min}(A)$. Inserting this inequality gives
	\[
	\|d\|_2\le L_T\bar\delta\Bigl(1+\frac{\bar b+\|v\|_2}{\sigma_{\min}\umax}\Bigr)=\alpha_1\bar\delta+\alpha_2\bar\delta\|v\|_2 .
	\]
	
	\emph{(a)} If $u\in\U$, \cref{lem:delta}(i) gives $\|\delta\|_2\le\bar\delta$ and thus $\|d\|_2\le L_T\bar\delta$.
	
	\emph{(b)} Let $R$ collect the rows of $DT$ that correspond to the last coordinate of each chain, $R=[\nabla\Lfh{r_i-1}h_i]_i$, so that $\|R\|_2\le L_T$, $\hat b=R\hat f$ and $A=R\hat g$. Then $\|\hat b\|_2\le L_T\bar f$ and $\|A\|_2\le L_T\bar{\hat g}$, and $\|v\|_2=\|\hat b+Au\|_2\le L_T\bar f+L_T\bar{\hat g}\,\bar u$ with $\bar u=\sqrt m\,\umax$.
	
	\emph{(c)} Because $z^{\rm d}$ is the Brunovsky state of the reference, $\hat x-T^{-1}(z^{\rm d})=T^{-1}(z)-T^{-1}(z^{\rm d})$, and $T^{-1}$ is $L_{T^{-1}}$-Lipschitz on $\Z$.
	
	\emph{Rössler model.} $T$ is linear, $DT=M$, the constants follow from the values in the previous subsection, and \eqref{eq:alphas} evaluates to \eqref{eq:alphas-ross}.
\end{proof}

\begin{remark}
	The bound \eqref{eq:thm1} contains three ingredients with distinct origins. The factor $L_T$ is the geometric price of the coordinate change. The ratio $\bar b/(\sigma_{\min}\umax)$ records how much of the actuator budget the linearizing term consumes: the larger the cancellation effort, the larger the input, and hence the larger the part of $\delta$ that scales with $u$. The slope $\alpha_2$ measures the sensitivity of the disturbance to the virtual input through the inverse of the decoupling matrix. When $\|v\|_2\le v_{\max}$ of \eqref{eq:vmax}, one finds $\alpha_1+\alpha_2v_{\max}=2L_T$, twice the direct bound of part~(a); we return to this point after \cref{cor:W}.
\end{remark}

\begin{table}[h]
	\centering
	\caption{Constants of the R\"ossler model entering \cref{thm:prop} (computed from \eqref{eq:A-ross} and $T(x)=Mx$).}
	\label{tab:consts}
	\begin{tabular}{lcc}
		\toprule
		Quantity & Symbol & Value\\
		\midrule
		Lipschitz constant of $T$ & $L_T$ & $1.4605$\\
		Lipschitz constant of $T^{-1}$ & $L_{T^{-1}}$ & $2.9247$\\
		Smallest singular value of $A$ & $\sigma_{\min}$ & $0.3424$\\
		Largest singular value of $A$ & $\sigma_{\max}$ & $1.4604$\\
		Input bound & $\umax$ & $1100$\\
		Slope coefficient & $\alpha_2$ & $3.878\times10^{-3}$\\
		Offset coefficient & $\alpha_1$ & $1.4605+3.878\times10^{-3}\,\bar b\approx147.0$\\
		Cancellation effort & $\bar b$ & $3.75\times10^{4}$\\
		\bottomrule
	\end{tabular}
\end{table}

\subsection{Disturbance set for the tube}

Let $\bar v$ be any bound on $\|v\|_2$ over the operating range. In the design, $\bar v=v_{\max}$ of \eqref{eq:vmax}; the a priori choice $\bar v=\bar v_{\rm a}$ of \cref{thm:prop}(b) is always valid.

\begin{corollary}[Disturbance set for tube MPC]\label{cor:W}
	With
	\begin{equation}\label{eq:wbar}
		\bar w:=(\alpha_1+\alpha_2\bar v)\,\bar\delta,
	\end{equation}
	the set
	\begin{equation}\label{eq:W}
		\W:=\{w\in\R^{2m}:\ \|w\|_2\le\bar w\}
	\end{equation}
	is nonempty, convex and compact, and contains $d(z,v)$ for every $z\in\Z$ and every $v$ with $\|v\|_2\le\bar v$ that can occur with $u\in\U$. It can therefore serve as the disturbance set in the tube-MPC design.
\end{corollary}

\begin{proof}
	$\W$ is a closed Euclidean ball of radius $\bar w>0$ in $\R^{2m}$, hence nonempty, convex and compact. By \eqref{eq:thm1}, $\|d(z,v)\|_2\le\bar w(v)\le\bar w$ whenever $\|v\|_2\le\bar v$, which is the inclusion.
\end{proof}

\begin{remark}
	Whenever $u\in\U$ in the closed loop (see \eqref{eq:usat}), part~(a) of \cref{thm:prop} supplies the smaller radius $\bar w_{\U}:=L_T\bar\delta\le\bar w$ when $\bar v=v_{\max}$, and the set $\{\|w\|_2\le\bar w_{\U}\}$ is also valid. The affine form \eqref{eq:thm1} is the one that remains valid without the a priori knowledge $u\in\U$, for example for candidate predictions in which the input constraint is relaxed, and it exposes the dependence on $\|v\|_2$. Either radius can be used in \Cref{sec:tube}; the case study should state which one is used. In what follows $\bar w$ denotes the radius of \cref{cor:W}, and all statements remain true with $\bar w_{\U}$ in its place.
\end{remark}

\section{Robust Tube-Based Predictive Control}\label{sec:tube}

\subsection{Exact ZOH discretization}

Let $\tau$ be the sampling period, and hold $\nu$ constant on $[k\tau,(k+1)\tau)$, $\nu(t)=\nu_k$. Integrating \eqref{eq:tracking} exactly,
\begin{equation}\label{eq:disc}
	\tilde z_{k+1}=A_{\rm d}\tilde z_k+B_{\rm d}\nu_k+w_k,\qquad A_{\rm d}=e^{A_{\rm c}\tau},\quad B_{\rm d}=\int_0^\tau e^{A_{\rm c}s}B_{\rm c}\,ds,
\end{equation}
and, since $A_{\rm c}$ is nilpotent,
\[
A_{\rm d}=I_2\otimes\begin{bmatrix}1&\tau\\0&1\end{bmatrix},\qquad B_{\rm d}=I_2\otimes\begin{bmatrix}\tau^2/2\\ \tau\end{bmatrix}.
\]
The discrete residual is $w_k=\int_0^\tau e^{A_{\rm c}(\tau-s)}\,d(k\tau+s)\,ds$. Here $e^{A_{\rm c}s}=I+A_{\rm c}s$ and $\|e^{A_{\rm c}s}\|_2\le1+s$, hence
\begin{equation}\label{eq:wk}
	\|w_k\|_2\le\gamma(\tau)\,\bar w,\qquad \gamma(\tau):=\tau+\tfrac12\tau^2,
\end{equation}
as long as $d$ obeys the bound of \cref{thm:prop} over the interval. The sampling period is $\tau=0.005$~s.

\subsection{Disturbance set}

The discrete-time disturbance set is
\begin{equation}\label{eq:Wtau}
	\W_{\rm d}:=\gamma(\tau)\,\W=\{w:\|w\|_2\le\bar w_\tau\},\qquad \bar w_\tau:=\gamma(\tau)\,\bar w,
\end{equation}
with $\W$ from \cref{cor:W}. For a polyhedral implementation, $\W_{\rm d}$ is replaced by the smallest box $\{\|w\|_\infty\le\bar w_\tau\}$ that contains it; this enlarges the set but preserves every result below.

\subsection{Ancillary controller and the tube set}

Let $Q\succ0$ and $R\succ0$ be weights and $P$ the solution of the discrete algebraic Riccati equation for $(A_{\rm d},B_{\rm d},Q,R)$; it defines the terminal cost and the terminal feedback $K_{\rm f}:=-(R+B_{\rm d}^\top PB_{\rm d})^{-1}B_{\rm d}^\top PA_{\rm d}$. The ancillary gain $K$ is obtained by pole placement on each channel $(A_{\rm d},B_{\rm d})$ of the discretized double integrator: for the continuous poles $s_1,s_2$ let $z_i=e^{s_i\tau}$, $\sigma=z_1+z_2$, $\pi=z_1z_2$, and
\begin{equation}\label{eq:K}
	K=\Bigl[\tfrac{\sigma-\pi-1}{\tau^2},\ \tfrac{\sigma+\pi-3}{2\tau}\Bigr],\qquad \Phi:=A_{\rm d}+B_{\rm d}K .
\end{equation}
With $s_1=-12$ and $s_2=-16$ this gives $K=[-179.09,\,-26.58]$, and $\Phi$ has the real eigenvalues $0.9418$ and $0.9231$.
The matrix $\Phi$ is Schur, so $\|\Phi^j\|_2\le c_\Phi\lambda_\Phi^{\,j}$ for some $c_\Phi\ge1$ and $\lambda_\Phi\in(0,1)$. The weights are $Q=\diag(Q_1,Q_2)=\diag(1000,100,1000,100)$ and $R=0.001\,I_2$.

The tube cross-section is an RPI set $\E$ for the deviation dynamics $e^+=\Phi e+w$, $w\in\W_{\rm d}$. We take the outer approximation of the minimal RPI set~\cite{Rakovic2005,Kolmanovsky1998}, $\E=(1-\lambda_{\rm s})^{-1}\bigoplus_{j=0}^{s-1}\Phi^j\W_{\rm d}$ with $\Phi^s\W_{\rm d}\subseteq\lambda_{\rm s}\W_{\rm d}$. Two properties are used. First, $\Phi\E\oplus\W_{\rm d}\subseteq\E$. Second, $\E\subseteq\rho_{\E}\,\mathbb B$ with the unit Euclidean ball $\mathbb B$ and
\begin{equation}\label{eq:rhoE}
	\rho_{\E}=c_{\E}\,\gamma(\tau)\,\bar w,\qquad c_{\E}:=\frac{c_\Phi}{(1-\lambda_\Phi)(1-\lambda_{\rm s})},
\end{equation}
where $c_{\E}$ depends only on $(A_{\rm d},B_{\rm d},K)$ and not on $\bar w$. In particular $\rho_\E\to0$ as $\bar\delta\to0$. In the implementation, the disturbance set is the box $|w_c|\le w_{\max,c}$ of each channel $c$, and $\E$ is the box in the modal coordinates $\eta=Me$ of $\Phi=T\Lambda M$, $\Lambda=\diag(\lambda_1,\lambda_2)$: $|\eta_i|\le c_i:=|MB_{\rm d}|_i\,w_{\max,c}/(1-|\lambda_i|)$, which is an exact RPI set for $\eta^+=\Lambda\eta+MB_{\rm d}w$. It gives the half-widths $(0.8215,\,11.26)$ in position and velocity for channel~1 and $(0.5281,\,7.239)$ for channel~2, and the tightening $\max_{\varepsilon\in\E}|K\varepsilon|=152.16$ and $97.82$, so that $|v_{\rm nom}|\le347.84$ and $402.18$ for $v_{\max}=500$.

\subsection{Constraint tightening}

Three constraint sets enter. The first keeps the response inside the region where the analysis holds. Let $\D:=\overline{\{z^{\rm d}(t):t\ge0\}}$, and let $\zeta:=\sup\{\|\tilde z\|_2:\tilde z\in\Z\ominus\D\}$. The tracking-coordinate state set is
\begin{equation}\label{eq:Zc}
	\Z_{\rm c}:=(\Z\ominus\D)\ominus\vartheta\,\mathbb B,\qquad \vartheta:=\tau\,(\zeta+\nu_{\max}+\bar w),
\end{equation}
which contains the origin in its interior by \cref{as:compact}; the margin $\vartheta$ accounts for the intersample motion of $\tilde z$. The second is the virtual input set $\V:=\{\nu:\|\nu\|_\infty\le\nu_{\max}/\sqrt m\}$, contained in the ball of radius $\nu_{\max}$ of \eqref{eq:vmax}. The tightened sets are
\begin{equation}\label{eq:tight}
	\bar\Z:=\Z_{\rm c}\ominus\E,\qquad \bar\V:=\V\ominus K\E .
\end{equation}
The tightening requires $\bar\Z$ and $\bar\V$ to have nonempty interior; otherwise $\bar\delta$ or $\tau$ must be reduced, which is a direct consequence of the size of $\E$ in \eqref{eq:rhoE}. For the R\"ossler model $T$ is linear, so a box $\X$ yields a polytope $\Z$ and all sets are computed by linear programming.

\subsection{Nominal optimal control problem}

Let $\ell(\bar z,\bar\nu)=\bar z^\top Q\bar z+\bar\nu^\top R\bar\nu$ and $V_{\rm f}(\bar z)=\bar z^\top P\bar z$. Let $\Z_{\rm f}\subseteq\bar\Z$ be a positively invariant set of $\bar z^+=(A_{\rm d}+B_{\rm d}K_{\rm f})\bar z$ with $K_{\rm f}\Z_{\rm f}\subseteq\bar\V$ and $0\in\operatorname{int}\Z_{\rm f}$; the maximal such set is a polytope. With the prediction horizon $N=20$, the problem at the tracking state $\tilde z_k$ is
\begin{equation}\label{eq:ocp}
	\begin{aligned}
		\mathbb P_N(\tilde z_k):\ V_N^\ast(\tilde z_k)=\min_{\bar z_0,\,\bar\nu_{0:N-1}}\ &\sum_{j=0}^{N-1}\ell(\bar z_j,\bar\nu_j)+V_{\rm f}(\bar z_N)\\
		\text{s.t. }\ &\tilde z_k-\bar z_0\in\E,\\
		&\bar z_{j+1}=A_{\rm d}\bar z_j+B_{\rm d}\bar\nu_j,\quad j=0,\dots,N-1,\\
		&\bar z_j\in\bar\Z,\quad \bar\nu_j\in\bar\V,\quad \bar z_N\in\Z_{\rm f}.
	\end{aligned}
\end{equation}
The initial nominal state $\bar z_0$ is a decision variable constrained to $\tilde z_k\ominus\E$. With polytopic sets the problem is a strictly convex quadratic program. Denote its minimizer by $(\bar z_0^\ast,\bar\nu_{0:N-1}^\ast)$. The applied virtual increment is the tube control
\begin{equation}\label{eq:kappa}
	\nu_k=\kappa(\tilde z_k):=\bar\nu_0^\ast+K\bigl(\tilde z_k-\bar z_0^\ast\bigr).
\end{equation}
Let $\mathcal F_N$ denote the set of $\tilde z$ for which \eqref{eq:ocp} is feasible.

\subsection{Recursive feasibility}

\begin{theorem}[Recursive feasibility]\label{thm:rf}
	Let \cref{as:smooth,as:obs,as:fl,as:compact} hold, let $\bar\delta$ satisfy \eqref{eq:lemma1}, and let $\E$, $\bar\Z$, $\bar\V$, $\Z_{\rm f}$ be as above with $\nu_{\max}>0$. If $\mathbb P_N(\tilde z_0)$ is feasible at the activation instant, then for the closed loop \eqref{eq:tracking}, \eqref{eq:kappa}, for every realization of $\delta$ obeying \eqref{eq:deltabar}: (i) $\mathbb P_N(\tilde z_k)$ is feasible for all $k\ge0$; (ii) $\tilde z(t)\in\Z\ominus\D$, hence $\hat x(t)\in\X$, for all $t\ge 150$~s; (iii) the unclamped input obeys $\|u(t)\|_2\le(\nu_{\max}+\bar\beta)/\sigma_{\min}$ of \eqref{eq:usat}; in particular $u(t)\in\U$, so the saturation $|u_i|\le1100$ is never active, whenever this bound does not exceed $\umax$, and otherwise its inactivity has to be verified on the closed loop.
\end{theorem}

\begin{proof}
	The proof uses induction on $k$ over the statement $\mathcal S_k$: $\mathbb P_N(\tilde z_k)$ is feasible.
	
	\emph{Step 1: admissibility of the interval, given $\mathcal S_k$.} Let $(\bar z^\ast_j,\bar\nu^\ast_j)$ be a feasible (optimal) solution and set $\nu_k=\kappa(\tilde z_k)$. With $e_k:=\tilde z_k-\bar z_0^\ast\in\E$, $\nu_k=\bar\nu^\ast_0+Ke_k\in\bar\V\oplus K\E\subseteq\V$, hence $\|\nu_k\|_2\le\nu_{\max}$ and $\|v\|_2\le\bar v_{\rm d}+\nu_{\max}=v_{\max}$ by \eqref{eq:vmax}; by \eqref{eq:u-law} and \eqref{eq:usat}, $\|u(t)\|_2\le(\nu_{\max}+\bar\beta)/\sigma_{\min}$ as long as $\hat x(t)\in\X$. Also $\tilde z_k=\bar z_0^\ast+e_k\in\bar\Z\oplus\E\subseteq\Z_{\rm c}$. Suppose the response leaves $\X$ during the interval, and let $t^\ast$ be the first exit time. On $[k\tau,t^\ast)$, \cref{thm:prop} and \cref{cor:W} give $\|d\|_2\le\bar w$, and \eqref{eq:tracking} yields $\|\dot{\tilde z}\|_2\le\|A_{\rm c}\|_2\zeta+\nu_{\max}+\bar w$, using $\|A_{\rm c}\|_2=\|B_{\rm c}\|_2=1$ and $\tilde z(t)\in\Z\ominus\D$ before exit. Thus $\|\tilde z(t)-\tilde z_k\|_2\le\vartheta$, and $\tilde z_k\in\Z_{\rm c}$ implies $\tilde z(t)\in\Z\ominus\D$, i.e.\ $\hat x(t)=T^{-1}(\tilde z(t)+z^{\rm d}(t))\in\X$ up to the boundary, contradicting the definition of $t^\ast$ by continuity. Therefore the response stays in $\X$, $u\in\U$, and \eqref{eq:wk} holds: $w_k\in\W_{\rm d}$.
	
	\emph{Step 2: the deviation stays in the tube.} With $\bar z^\ast_1=A_{\rm d}\bar z_0^\ast+B_{\rm d}\bar\nu_0^\ast$, subtracting from \eqref{eq:disc} gives
	\[
	\tilde z_{k+1}-\bar z_1^\ast=\Phi e_k+w_k\in\Phi\E\oplus\W_{\rm d}\subseteq\E .
	\]
	
	\emph{Step 3: a feasible candidate at $k+1$.} Take $\bar z_0^{+}=\bar z_1^\ast$, $\bar\nu^{+}_j=\bar\nu^\ast_{j+1}$ for $j\le N-2$, and $\bar\nu^{+}_{N-1}=K_{\rm f}\bar z_N^\ast$. The initial constraint holds by Step~2. For $j\le N-2$ the shifted states and inputs lie in $\bar\Z$ and $\bar\V$ by feasibility at $k$. Since $\bar z_N^\ast\in\Z_{\rm f}$ and $K_{\rm f}\Z_{\rm f}\subseteq\bar\V$, the last input is admissible, and positive invariance places $\bar z_N^{+}=(A_{\rm d}+B_{\rm d}K_{\rm f})\bar z_N^\ast\in\Z_{\rm f}\subseteq\bar\Z$. Hence $\mathcal S_{k+1}$ holds. Because $\mathcal S_0$ is assumed, the induction closes.
	
	Claims (ii) and (iii) were established in Step~1 for every $k$.
\end{proof}

\begin{remark}
	The theorem places no requirement on the sign or size of the identification error beyond $\bar\delta$; when the tightened sets are empty for the value of $\bar\delta$ at hand, the optimization is infeasible from the start, and the remedy is to reduce $\tau$, improve the identification, or enlarge $\X$. Feasibility of $\mathbb P_N(\tilde z_0)$ at the activation instant $t=150$~s holds when $\tilde z_0\in\mathcal F_N$.
\end{remark}

\subsection{Input-to-state stability}

Let $\lambda_Q:=\lambda_{\min}(Q)$ and recall $V_N^\ast$ from \eqref{eq:ocp}. Assume in addition that $\bar\Z$ and $\bar\V$ contain the origin in their interior, so that $0\in\operatorname{int}\Z_{\rm f}$ and a ball $r\mathbb B$ of radius $r>0$ lies in the set of nominal states for which \eqref{eq:ocp} with $\bar z_0=\tilde z_k$ is feasible.

\begin{theorem}[Input-to-state stability]\label{thm:iss}
	Under the hypotheses of \cref{thm:rf}, there are constants $c_2>0$ and $\mu\in(0,1]$ such that for every $\tilde z_0\in\mathcal F_N$ and every realization of $\delta$ obeying \eqref{eq:deltabar},
	\begin{equation}\label{eq:iss}
		\|\tilde z_k\|_2\le\beta(\|\tilde z_0\|_2,k)+\gamma_{\rm iss}\,\bar\delta,\qquad \beta(s,k)=\sqrt{c_2/\lambda_Q}\,(1-\mu)^{k/2}s,
	\end{equation}
	with
	\begin{equation}\label{eq:gain}
		\gamma_{\rm iss}=\Bigl(1+\sqrt{2c_2/\lambda_Q}\Bigr)\,c_{\E}\,\gamma(\tau)\,(\alpha_1+\alpha_2\bar v).
	\end{equation}
	In particular the ISS gain is linear in $\bar\delta$ and vanishes as $\bar\delta\to0$, and the synchronization errors satisfy $|e_1(k\tau)|,|e_2(k\tau)|\le\|\tilde z_k\|_2$.
\end{theorem}

\begin{proof}
	\emph{Value-function bounds.} Write $V:=V_N^\ast$. The stage cost gives the lower bound $V(\tilde z)\ge\lambda_Q\|\bar z_0^\ast\|_2^2$. For the upper bound, $V$ is bounded on the compact set $\mathcal F_N$, say by $V_{\max}$. For $\|\tilde z\|_2\le r$ the choice $\bar z_0=\tilde z$ with the unconstrained LQR sequence is feasible and gives $V(\tilde z)\le\lambda_{\max}(P)\|\tilde z\|_2^2$. For $\|\tilde z\|_2>r$, $V\le V_{\max}\le(V_{\max}/r^2)\|\tilde z\|_2^2$. With $c_2:=\max\{\lambda_{\max}(P),V_{\max}/r^2\}$,
	\[
	V(\tilde z)\le c_2\|\tilde z\|_2^2\quad\text{on }\mathcal F_N .
	\]
	
	\emph{Decrease.} By \cref{thm:rf}, the candidate of Step~3 is feasible at $k+1$. Its cost is the optimal cost at $k$ minus $\ell(\bar z_0^\ast,\bar\nu_0^\ast)$ plus the terminal terms $-V_{\rm f}(\bar z_N^\ast)+\ell(\bar z_N^\ast,K_{\rm f}\bar z_N^\ast)+V_{\rm f}((A_{\rm d}+B_{\rm d}K_{\rm f})\bar z_N^\ast)$, and the latter sum is nonpositive by the Riccati equation for $K_{\rm f}=K$. Therefore
	\[
	V(\tilde z_{k+1})\le V(\tilde z_k)-\ell(\bar z_0^\ast,\bar\nu_0^\ast)\le V(\tilde z_k)-\lambda_Q s_k^2,\qquad s_k:=\|\bar z_0^\ast(\tilde z_k)\|_2 .
	\]
	
	\emph{Recursion.} Since $\tilde z_k-\bar z_0^\ast\in\E\subseteq\rho_\E\mathbb B$, we have $\|\tilde z_k\|_2\le s_k+\rho_\E$, hence $\|\tilde z_k\|_2^2\le2s_k^2+2\rho_\E^2$ and $V(\tilde z_k)\le 2c_2s_k^2+2c_2\rho_\E^2$. Solving for $s_k^2$ and inserting in the decrease inequality,
	\[
	V(\tilde z_{k+1})\le(1-\mu)V(\tilde z_k)+2\mu c_2\rho_\E^2,\qquad \mu:=\min\{1,\lambda_Q/(2c_2)\}.
	\]
	Iterating and summing the geometric series, $V(\tilde z_k)\le(1-\mu)^kV(\tilde z_0)+2c_2\rho_\E^2\le(1-\mu)^kc_2\|\tilde z_0\|_2^2+2c_2\rho_\E^2$.
	
	\emph{Conversion.} From $\|\tilde z_k\|_2\le s_k+\rho_\E\le\sqrt{V(\tilde z_k)/\lambda_Q}+\rho_\E$ and $\sqrt{a+b}\le\sqrt a+\sqrt b$,
	\[
	\|\tilde z_k\|_2\le\sqrt{c_2/\lambda_Q}\,(1-\mu)^{k/2}\|\tilde z_0\|_2+\bigl(1+\sqrt{2c_2/\lambda_Q}\bigr)\rho_\E .
	\]
	Insertion of $\rho_\E=c_\E\gamma(\tau)\bar w=c_\E\gamma(\tau)(\alpha_1+\alpha_2\bar v)\bar\delta$ from \eqref{eq:rhoE} yields \eqref{eq:iss}--\eqref{eq:gain}. The bound on $e_1,e_2$ holds because $e_1=\tilde z_1$ and $e_2=\tilde z_3$.
\end{proof}

\begin{remark}
	The bound holds uniformly over all realizations of $\delta$ with $\sup\|\delta\|_2\le\bar\delta$, where $\bar\delta$ is the level for which the tube is designed. The gain \eqref{eq:gain} multiplies $\bar\delta$ by a factor that depends on the geometry of the coordinate change ($L_T$ through $\alpha_1,\alpha_2$), on the actuator budget ($\bar b/\sigma_{\min}\umax$), on the discretization ($\gamma(\tau)$) and on the LQR design ($c_\E$, $c_2$, $\lambda_Q$). Improving the identification lowers $\bar\delta$ and the ultimate bound with it; when $\bar\delta=0$ the tube collapses to $\{0\}$, the nominal predictive controller of \eqref{eq:ocp} is recovered, and \eqref{eq:iss} gives exponential convergence of $\tilde z_k$. The inequality is an ISS estimate in the sense of tube-based MPC~\cite{Limon2010,Mayne2000}, and is equivalent to ISS with respect to the disturbance sequence $(w_k)$, whose size is fixed by $\bar w$.
\end{remark}

\subsection{Online implementation}

\Cref{alg:online} lists the computations at each sampling instant. The offline stage evaluates \eqref{eq:A-ross}--\eqref{eq:sigmin}, $\bar b$, $\bar\delta$ (from \Cref{lem:delta}), $\bar w$, $\W_{\rm d}$, $K$, $P$, $\E$, the tightened sets and $\Z_{\rm f}$. The online stage reads the state, evaluates $T$ and $A^\dagger$, solves one QP and applies the physical control.

\begin{algorithm}[h]
	\caption{Online implementation of the three-layer scheme}
	\label{alg:online}
	\begin{algorithmic}[1]
		\Require Identified $(\hat f,\hat g)$, $T$, $A$, $\hat b$, gain $K$, sets $\E,\bar\Z,\bar\V,\Z_{\rm f}$, weights $Q,R,P$, horizon $N$, period $\tau$, activation time $t_{\rm a}=150$~s
		\For{$k=0,1,2,\dots$ at $t_k=k\tau$}
		\State read $\hat x_k$ and drive state $x_k$; form $z^{\rm d}_k$ and $v_{\rm d}(t)$ from \eqref{eq:zd}
		\If{$t_k<t_{\rm a}$}
		\State $u\gets 0$ and \textbf{continue}
		\EndIf
		\State evaluate $z_k=T(\hat x_k)$ and $\tilde z_k=z_k-z^{\rm d}_k$
		\State\label{ln:qp} solve the tightened QP $\mathbb P_N(\tilde z_k)$ \eqref{eq:ocp} for $(\bar z_0^\ast,\bar\nu^\ast_{0:N-1})$
		\State $\nu_k\gets\bar\nu_0^\ast+K(\tilde z_k-\bar z_0^\ast)$
		\For{$t\in[t_k,t_{k+1})$}
		\State evaluate $A^\dagger(\hat x(t))$ and $\hat b(\hat x(t))$
		\State\label{ln:apply} apply $u(t)=\operatorname{sat}_{\umax}\!\bigl(A^\dagger(\hat x(t))\bigl(-\hat b(\hat x(t))+v_{\rm d}(t)+\nu_k\bigr)\bigr)$ \Comment{redundant clamp}
		\EndFor
		\EndFor
	\end{algorithmic}
\end{algorithm}

By \cref{thm:rf}, the QP in line~\ref{ln:qp} is feasible at every instant after activation and the input in line~\ref{ln:apply} lies in $\U$ before clipping; here $\operatorname{sat}_{\umax}$ denotes componentwise clipping to $[-\umax,\umax]$, so the clamp is a redundant hardware safeguard and no saturation logic is needed in the nominal operation of the loop.

\section{Case Study and Simulation Results}\label{sec:sim}

We now instantiate the pipeline of \cref{sec:sindy,sec:fl,sec:tube} on the coupled R\"ossler pair \eqref{eq:drive}--\eqref{eq:response}. The identification stage is validated first, which fixes $\bar\delta$; the closed-loop scheme is then run from the activation time $t=150$~s to the horizon $t_{\rm f}=250$~s.

\subsection{Data acquisition and library}
\begin{figure}[h]
	\centering
	\includegraphics[width=\columnwidth]{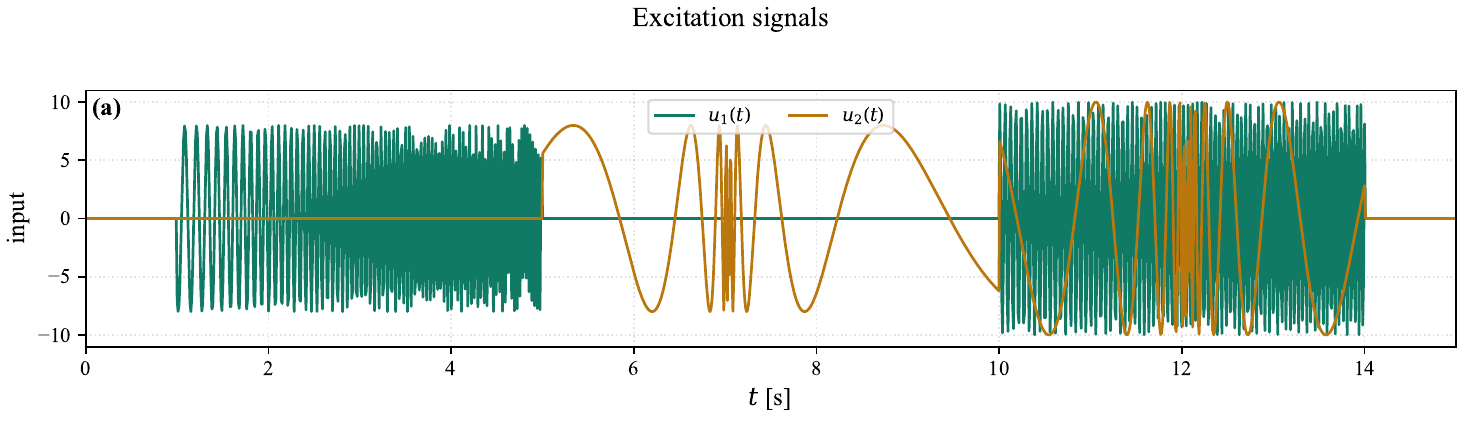}
	\caption{%
		Excitation signals applied to the plant for data collection.
	}
	\label{fig:inputs}
\end{figure}

Training data are generated by exciting the response system \eqref{eq:response} in open loop from the initial condition $X_0=[4,\,1,\,0.01,\,3]^\top$ on $t\in[0,15]$~s, with $N_s=1500$ uniformly spaced samples (step $15/1499\approx0.01$~s); the derivatives $\dot x_j$ are evaluated from the model equations at the sampled pairs $(\hat x_j,u_j)$, and the trajectory is integrated with the input interpolated linearly between samples. The excitation consists of chirp bursts, shown in \cref{fig:inputs}: $u_1=8\,c_{\rm lin}(t;3)$ on $[1,5]$~s and $u_1=10\,c_{\rm lin}(t;12)$ on $[10,14]$~s, and $u_2=8\,c_{\rm hyp}(t;7)$ on $[5,10]$~s and $u_2=10\,c_{\rm hyp}(t;12)$ on $[10,14]$~s, with both inputs zero elsewhere. Here $c_{\rm lin}(t;t_1)$ and $c_{\rm hyp}(t;t_1)$ denote the linear and hyperbolic chirps of \texttt{scipy.signal.chirp} with $f_0=0.1$~Hz, $f_1=20$~Hz and target time $t_1$, evaluated at the absolute time $t$. As the sweeps are not restarted in each window, the instantaneous frequency of the first linear burst runs from about $6.7$ to $33$~Hz and the hyperbolic bursts become broadband near their singularities at $t\approx7.04$~s and $t\approx12.06$~s; the two inputs are linearly independent, which makes $\hat g$ identifiable. The library is the degree-two monomial set described in \cref{sec:library}, and the coefficients $\hat\Xi$ are obtained by the STLSQ procedure of \cref{sec:stlsq} with threshold $\lambda=0.03$ and ridge parameter $\alpha=0.04$. This reproduces the nominal model \eqref{eq:sindy} to three decimals, consistent with $\bar\eta_f=\bar\eta_g=0$ for this plant.

\subsection{Validation of the identified model}\label{sec:validation}
\begin{figure}[h]
	\centering
	\includegraphics[width=\textwidth]{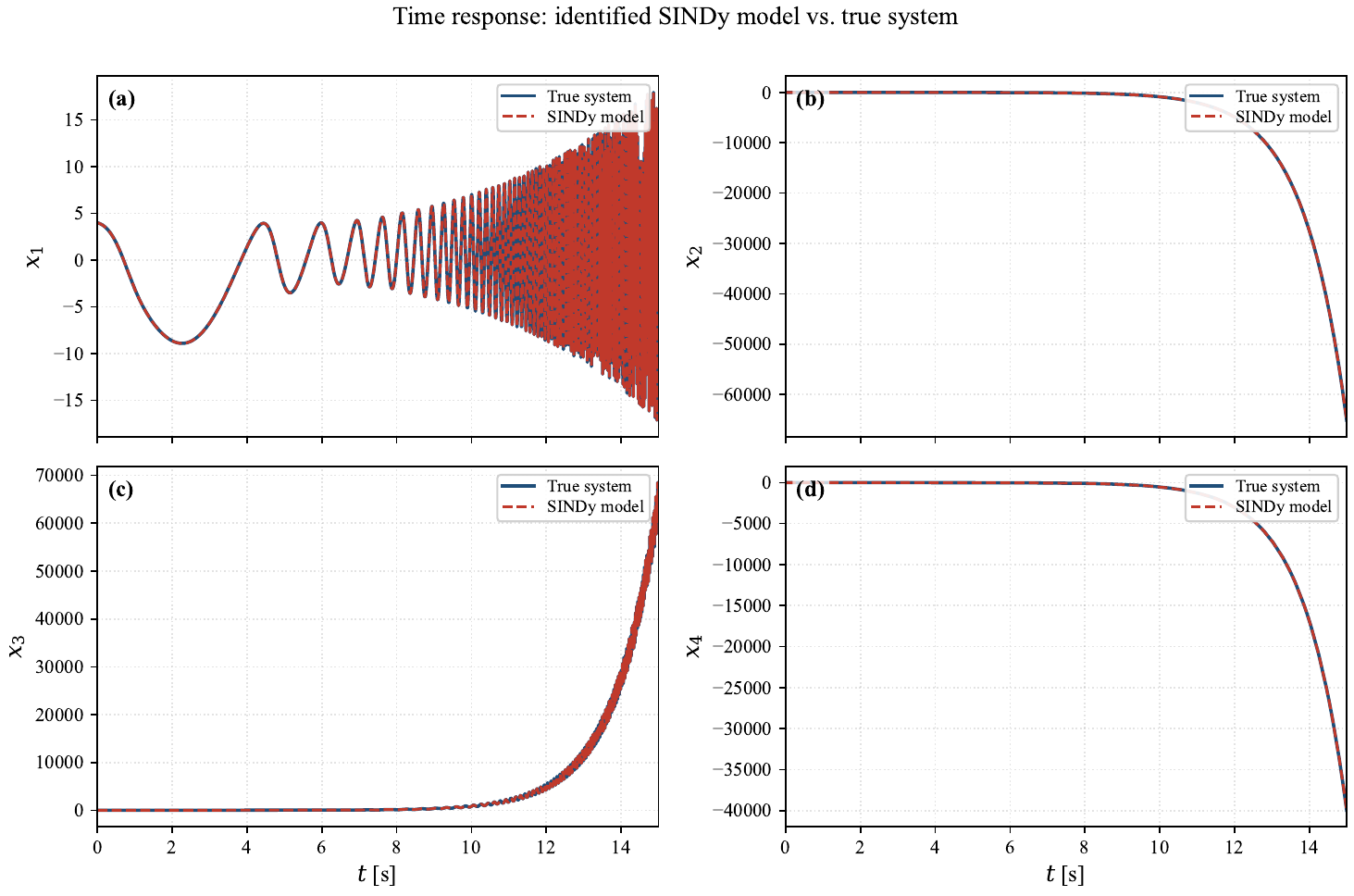}
	\caption{%
		Time-domain validation of the SINDy-identified model over a
		horizon of $T=15$\,s under the excitation signals of
		Fig.~\ref{fig:inputs}. Panels~(a)--(d) correspond to states
		$x_1$--$x_4$, respectively. In each panel, the solid blue
		line is the response of the true system and the dashed red
		line is the response of the SINDy model. 
	}
	\label{fig:timeseries}
\end{figure}

\begin{figure*}[h]
	\centering
	\includegraphics[width=\textwidth]{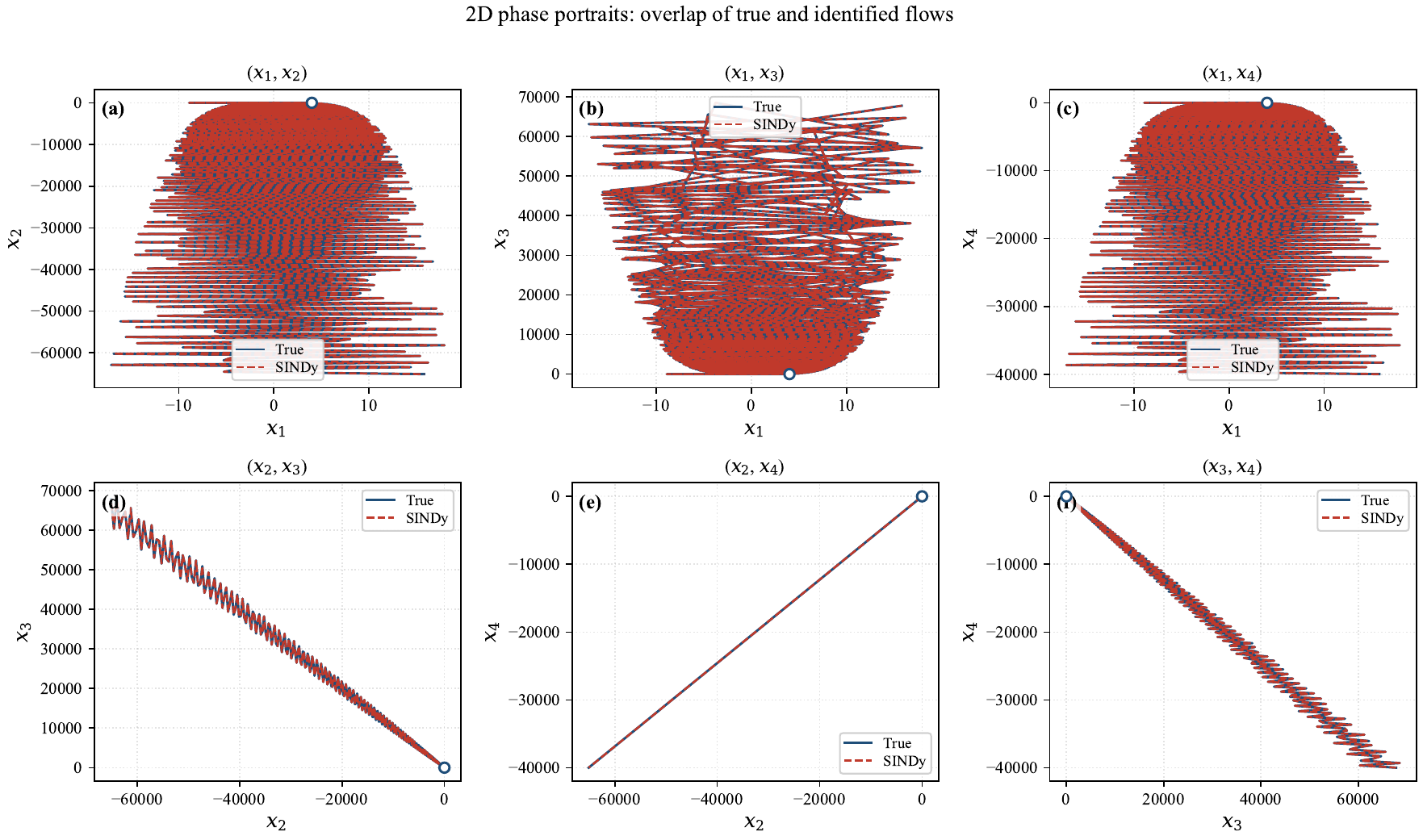}
	\caption{%
		Two-dimensional phase portraits of the true and identified
		flows. Panels~(a)--(f) correspond to the six coordinate pairs
		$(x_1,x_2)$, $(x_1,x_3)$, $(x_1,x_4)$, $(x_2,x_3)$,
		$(x_2,x_4)$ and $(x_3,x_4)$. In each panel, the solid blue
		curve is the true trajectory, the dashed red curve is the
		SINDy-identified one, and the white circle marks the initial
		condition. 
	}
	\label{fig:phase}
\end{figure*}

The model \eqref{eq:sindy} is validated open-loop against the true system over a horizon of $15$~s under the excitation of \cref{fig:inputs}, disjoint from the identification record. \Cref{fig:timeseries} compares the four state trajectories; the identified and true responses overlap without visible drift over the validation window, and \cref{fig:phase} confirms this in the six coordinate planes, which indicates that the residual $\delta(x,0)$ does not concentrate along a particular direction of $\X$.

\begin{figure}[h]
	\centering
	\includegraphics[width=\columnwidth]{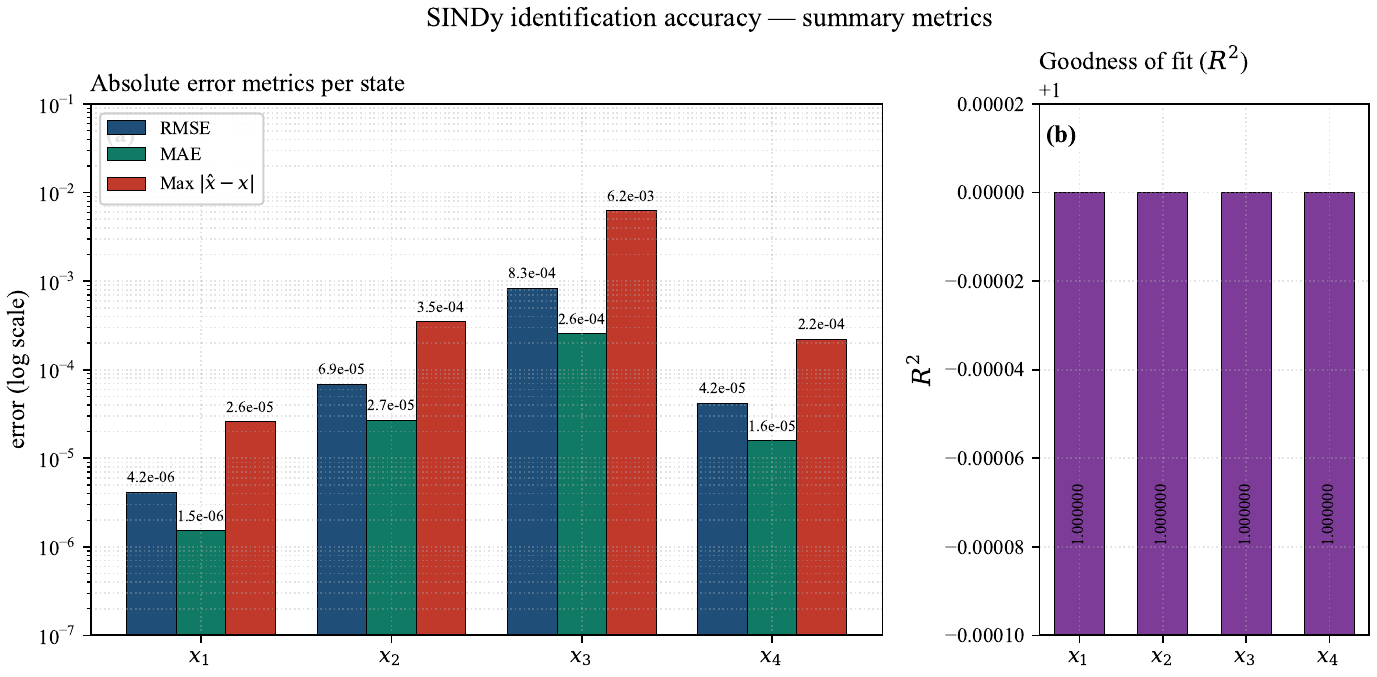}
	\caption{%
		Quantitative accuracy of the SINDy identification.
		Panel~(a) reports the absolute error metrics (RMSE, MAE, and
		maximum pointwise error) for each state on a logarithmic
		scale. Panel~(b) reports the coefficient of determination
		$R^2$ for each state; all values are equal to $1.0000$
		within four decimal places.%
	}
	\label{fig:metrics}
\end{figure}

\Cref{fig:metrics} reports the pointwise RMSE, mean absolute error and maximum error per state (panel a) and $R^2$ per state (panel b). All four states attain $R^2=1.0000$, and the RMSE ranges from $4.15\times10^{-6}$ to $8.33\times10^{-4}$, the largest value on $x_3$ because of the bilinear term $x_1x_3$ in $\hat f_3$. These are the per-state values collected in $\varepsilon_{\rm rmse}$ of \cref{sec:stlsq}.

\paragraph{Calibration of $\bar\delta$.} With $\|\varepsilon_{\rm rmse}\|_2\le2\times8.33\times10^{-4}$ from \cref{fig:metrics} and margin $\chi=2.7$, the calibrated bound of \eqref{eq:calib} used in the remainder of this section is $\bar\delta=\chi\cdot\rho_{\rm val}=2.7\times6.7\times10^{-4}\approx1.8\times10^{-3}$, where $\rho_{\rm val}=\max_j\|r_j^{\rm val}\|_2$ is evaluated along the closed-loop states and applied inputs of the simulation below. This is the value entered in \eqref{eq:wbar} to obtain $\bar w$ and, through \eqref{eq:Wtau}, $\bar w_\tau$.

\subsection{Closed-loop configuration}
The sampling period is $\tau=0.005$~s, the LQR weights are $Q=\diag(1000,100,1000,100)$ and $R=0.001\,I_2$, and the prediction horizon is $N=20$, chosen so that $\tilde z_0=\tilde z(150\,\mathrm s)\in\mathcal F_N$, as required by \cref{thm:rf}. The drive is initialized at $x(0)=[-50,\,-15,\,70,\,35]^\top$ and the response at $\hat x(0)=[-45,\,-10,\,68,\,34]^\top$; both evolve in open loop until $t=150$~s, when the scheme of \cref{alg:online} is switched on, and the simulation runs to $t_{\rm f}=250$~s. The input disturbance has amplitude $d_{\rm amp}=5$ with zero-order-hold margin $w_{\rm zoh}=11$, so that $w_{\max}=(21,\,13.5)$ by \eqref{eq:wmax}. The margin $w_{\rm zoh}$ is set from the transient: with $w_{\rm zoh}=2$ the effective disturbance observed during the first seconds after activation, $\max|w|=(21.1,\,11.5)$, exceeded $w_{\max}$ and the tube was violated in $5$ and $4$ of $20000$ steps, whereas $w_{\rm zoh}=11$ gives no violation. The identification error is negligible next to this bound: $\alpha_1=147.0$, $\bar w=(\alpha_1+\alpha_2v_{\max})\bar\delta=0.29$ and $\bar w_{\U}=L_T\bar\delta=2.6\times10^{-3}$, below $1.4\%$ of $w_{\max,1}$. The virtual input is constrained by the tightened set $\bar\V$ of \eqref{eq:tight} with $\nu_{\max}=500\sqrt2$; the saturation $|u_i|\le1100$ is not enforced by the QP but monitored, with the hardware clamp of line~\ref{ln:apply} of \cref{alg:online} as a safeguard.

\subsection{State trajectories and tracking performance}
\begin{figure}[!t]
	\centering
	\includegraphics[width=\linewidth]{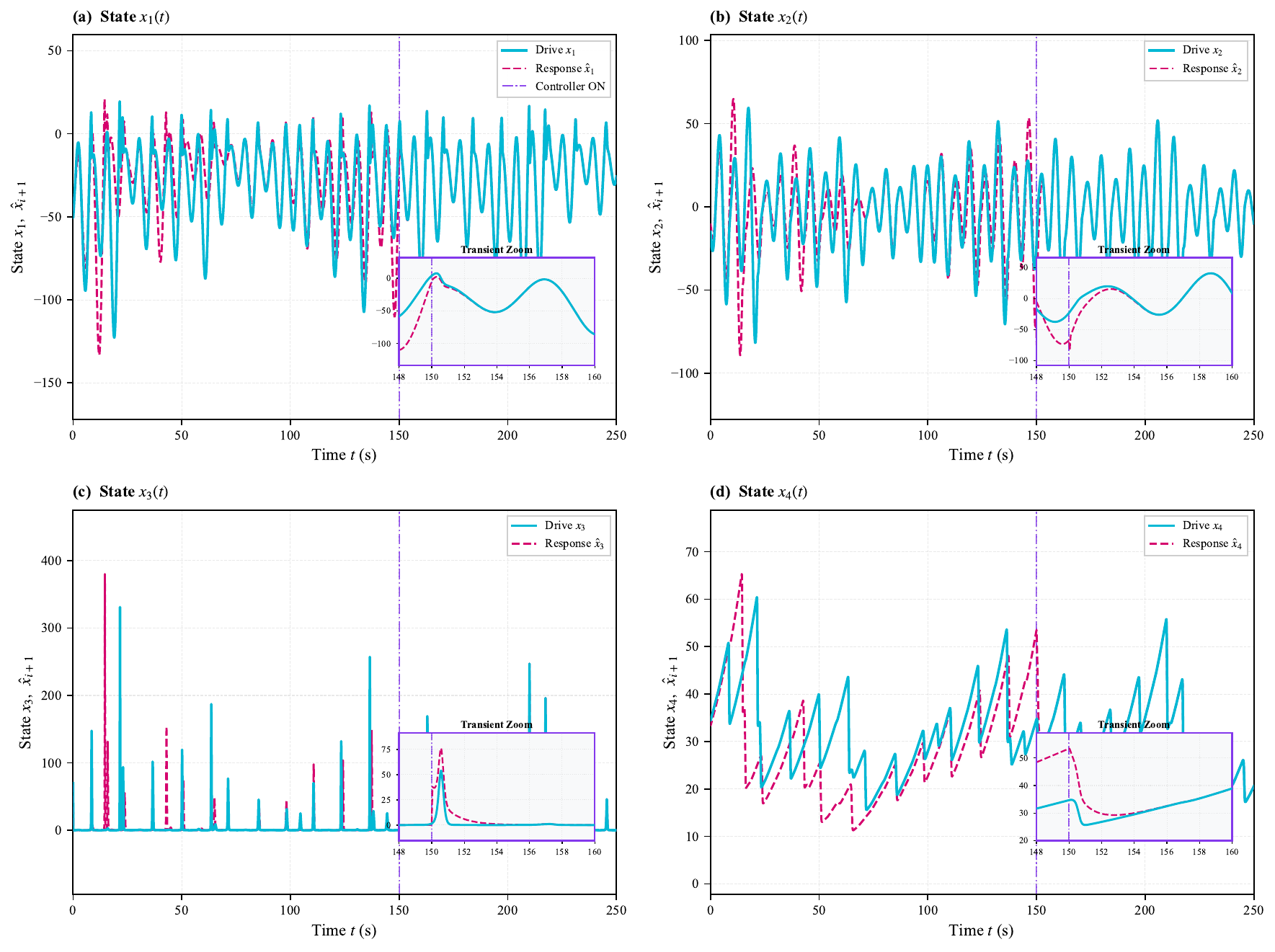}
	\caption{State trajectories of the drive system and response system with invariant tube bounds activated at $t = 150\,\text{s}$.}
	\label{fig:states}
\end{figure}

\Cref{fig:states} shows the four drive states $x_i(t)$ against the response states $\hat x_i(t)$ over $[0,250]$~s. Before activation the two systems diverge as expected of coupled chaotic oscillators with different initial conditions; from $t=150$~s the response is steered by \eqref{eq:kappa} and the trajectories lock onto each other, consistent with \cref{thm:iss} once $\tilde z_k$ has entered a neighbourhood of the ball $\Zt$ defined in \cref{sec:errors}.

\subsection{Synchronization error and tube bound}\label{sec:errors}
\begin{figure}[h]
	\centering
	\includegraphics[width=\linewidth]{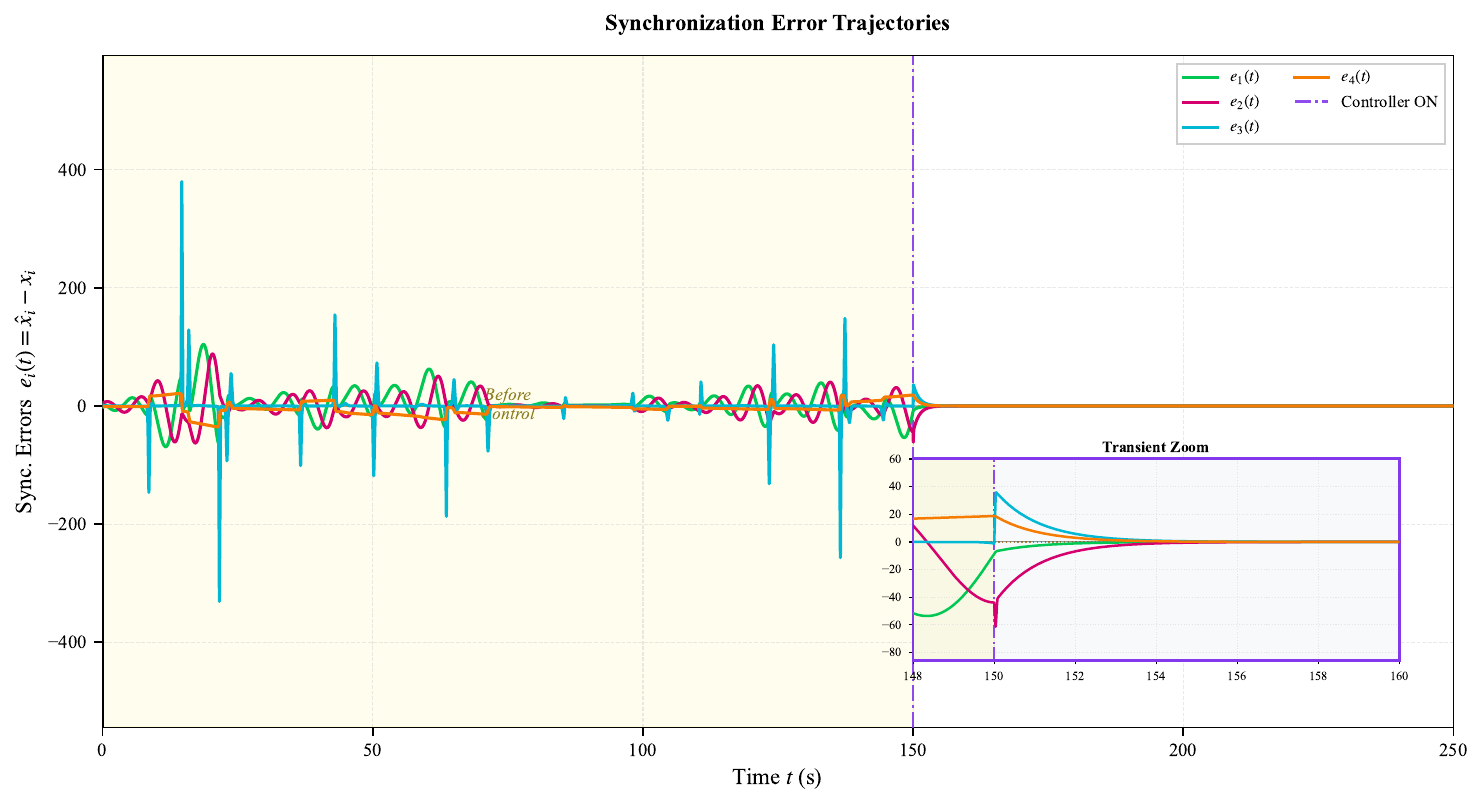}
	\caption{Synchronization error trajectories bounded within the robust invariant regions.}
	\label{fig:errors}
\end{figure}

\Cref{fig:errors} reports $\tilde z(t)=(e_1,\dot e_1,e_2,\dot e_2)$ together with the constant bound $\rho_{\rm t}:=\gamma_{\rm iss}\bar\delta$ of \cref{thm:iss}, with $\gamma_{\rm iss}$ from \eqref{eq:gain}; we write $\Zt:=\{\tilde z:\|\tilde z\|_2\le\rho_{\rm t}\}$ for the corresponding ball. After the transient following activation, $\tilde z_k$ enters and stays within the ball $\Zt$, and $e_1,e_2$ remain inside the interval $[-\rho_{\rm t},\rho_{\rm t}]$ predicted by \eqref{eq:iss}, without requiring $\bar\delta$ to be driven to zero. This is the behaviour \cref{thm:iss} certifies for a lumped uncertainty of the size reported in \cref{fig:metrics}.

\subsection{Control inputs and saturation}
\begin{figure}[h]
	\centering
	\includegraphics[width=\linewidth]{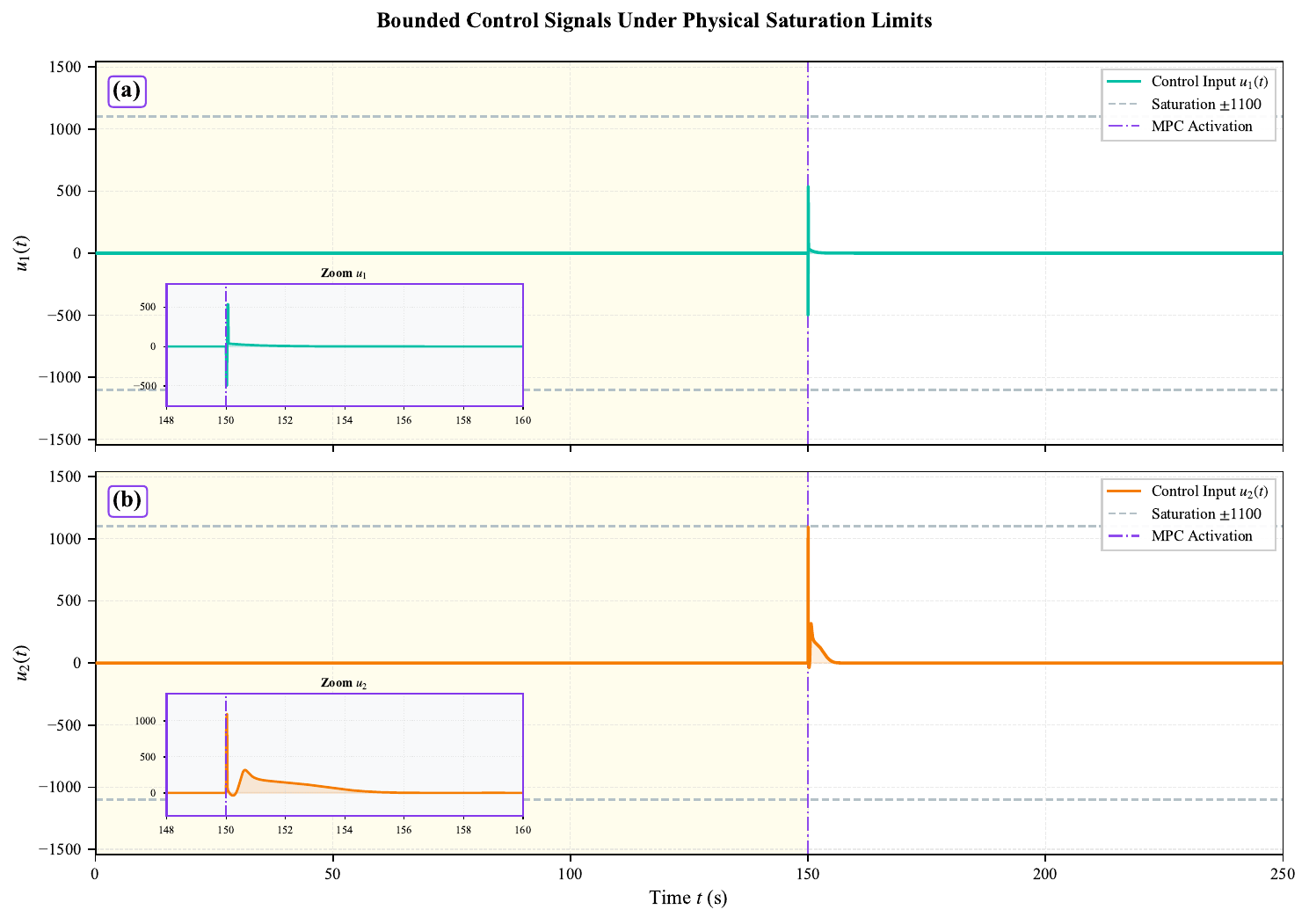}
	\caption{Bounded physical control input signals $u_1(t)$ and $u_2(t)$ satisfying actuator constraints.}
	\label{fig:controls}
\end{figure}

\Cref{fig:controls} shows $u_1(t)$ and $u_2(t)$ . The signals stay within $|u_i|\le1100$ throughout the transient that follows activation, with maxima $|u_1|=344.3$ and $|u_2|=769.7$, and the largest value reaches $70\%$ of the limit, so the hardware saturation of \cref{alg:online} is not triggered under these initial conditions; the worst-case bound \eqref{eq:usat} does not certify this, and the inactivity is verified on the closed loop.

\subsection{Three-dimensional phase-space view}
\begin{figure}[h]
	\centering
	\includegraphics[width=0.8\linewidth]{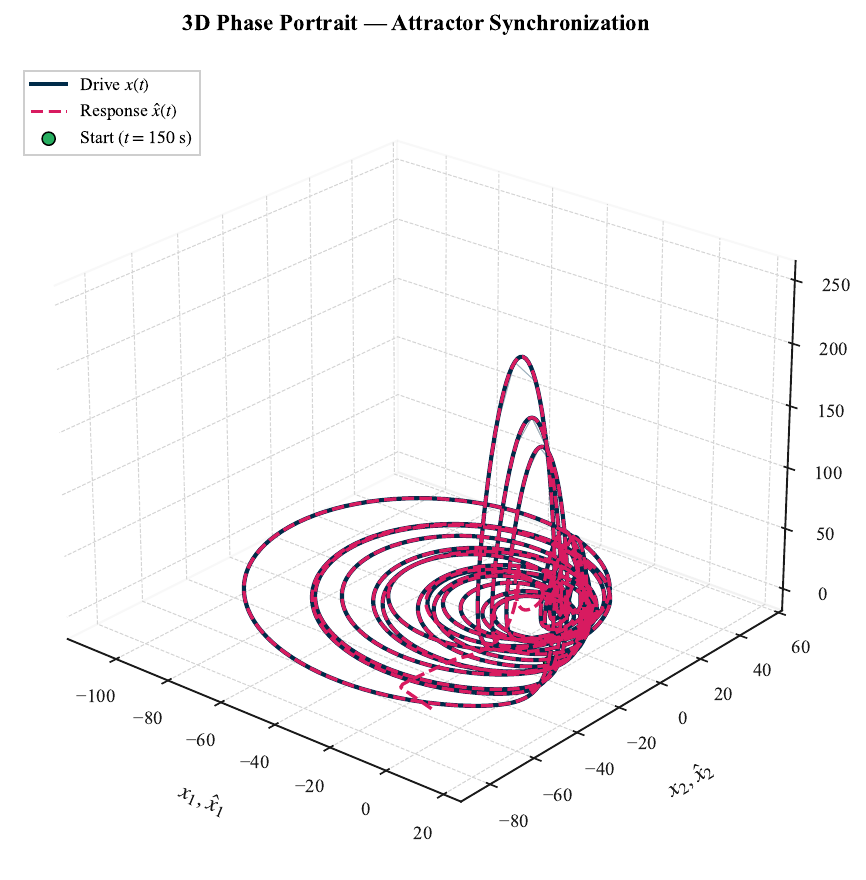}
	\caption{Three-dimensional phase space trajectory .}
	\label{fig:phase3d}
\end{figure}

\Cref{fig:phase3d} projects the closed loop onto $(x_1,x_2,x_3)$. The response trajectory departs from the drive attractor before activation and, once inside the tube, tracks it within a thin envelope, which is the geometric counterpart of the ball $\Zt\subset\R^4$ pulled back through $T^{-1}$.

\subsection{Discussion}
Three observations tie the simulation back to \cref{sec:sindy,sec:fl,sec:tube}. First, the identification residual reported in \cref{fig:metrics} is small enough that $\rho\approx0$ in \cref{lem:delta}, so the propagated disturbance of \cref{thm:prop} is dominated by the drift term $\alpha_1\bar\delta$ rather than by $\|v\|_2$, which is visible in the near-constant width of the tube in \cref{fig:errors}. Second, the tightening \eqref{eq:Zc}--\eqref{eq:tight} is conservative enough that no infeasibility was encountered over the horizon, in line with \cref{thm:rf}. Third, the saturation margin in \cref{fig:controls} indicates room for a more aggressive LQR weighting than $Q=\diag(1000,100,1000,100)$, $R=0.001\,I_2$ without leaving $\U$; the choice of the weights is a design degree of freedom that we do not explore here.

\section{Declarations}

\subsection{Funding}
This research received no specific grant from any funding agency in the public, commercial, or not-for-profit sectors.

\subsection{Author Contribution}
\begin{itemize}
	\item \textbf{Shahin Razani:} Conceptualization, Methodology, Investigation, Writing - Original Draft.
\end{itemize}

\subsection{Availability of Data and Materials}
The datasets generated during and/or analysed during the current study are available from the corresponding author on reasonable request.

\subsection{Competing Interests}
The author declares no financial or non-financial interests that are directly or indirectly related to the work submitted for publication. No competing interests arose during the last three years of conducting and preparing this research.
\section{Conclusion}\label{sec:conclusion}

This paper addresses a gap that appears whenever a model learned from data is used inside a constrained controller: the identification error is usually small, but its effect on the closed loop is rarely quantified. We collected every discrepancy between the identified and the true response system into one lumped uncertainty, bounded it on the region where the loop operates, and followed it through input--output feedback linearization, where it becomes an additive disturbance on a chain of integrators with a gain that we computed in closed form. That disturbance, together with a bounded input disturbance and a zero-order-hold margin, defines the set on which a tube-based predictive controller is built. For this controller we proved recursive feasibility under the tightened constraints and input-to-state stability of the tracking error, with an asymptotic gain that is linear in the identification bound.

On a hyperchaotic R\"ossler drive--response pair, $1500$ samples were enough for sparse regression to recover the eleven nonzero coefficients of the response model with an error below $10^{-4}$, which gave $\bar\delta=1.8\times10^{-3}$. With the input limited to $\pm1100$ and a bounded input disturbance, the controller kept the synchronization error inside the tube during all $20000$ controlled steps, without a failure of the quadratic program and with a peak input at $70\%$ of the limit. Two observations from the study matter to us as much as these numbers. First, the identification error enters the disturbance set only marginally: the radius it contributes is below $1.4\%$ of the per-channel bound fixed by the input disturbance and the zero-order-hold margin, so for a well-identified sparse model the tube is governed by the exogenous terms. Second, the worst-case bound on the physical input does not close for this plant, because the cancellation term of the linearization is much larger than the margin to the actuator limit. The absence of saturation was therefore verified on the closed loop and not proved, and a sharper certificate remains the main open theoretical question.

Several limitations point to the next steps. The identification data were free of measurement noise, the derivatives were exact, and the library contained the true structure; measurement noise, derivative estimation and model misspecification will enlarge $\bar\delta$ and will test the calibration procedure of \Cref{sec:delta}. The zero-order-hold margin was set from the transient observed in simulation, and an analytic bound that depends on the tracking error would remove this tuning step. Finally, the guarantees were examined on one plant, and extending the study to other chaotic systems, to larger ensembles of disturbance realizations, and to a hardware implementation is the natural continuation of this work.

\bibliography{DataDrivenFeedBack.bib}
\end{document}